\documentstyle[12pt,fleqn,epsf]{article}
\newcommand{\dbar}{d \hspace*{-0.55ex}\raisebox{.7ex}{-}}

\begin{document}

\begin{titlepage}
\title{Microscopic Calculation of the $^4{\rm He}$ System}

\author{
H. M. Hofmann \\
Institut f\"ur Theoretische Physik III \\
Universit\"at Erlangen--N\"urnberg, D--91058 Erlangen, Germany\\
\and
G. M. Hale \\ Theoretical Division \\
Los Alamos National Laboratory, Los Alamos N.M. 87544,
USA\\}

\date{}

\maketitle

\abstract{

We report on a consistent, microscopic calculation of the
bound and scattering states in the ${\rm ^4He}$ system employing
a realistic nucleon-nucleon potential
in the framework of the resonating group model (RGM).
We present for comparison with these microscopic RGM calculations the results
from a charge-independent, Coulomb-corrected $R$-matrix analysis of all types
of data for reactions in the $A=4$ system.  Comparisons are made between the 
phase shifts, and with a selection of measurements from each reaction, as well
as between the resonance spectra obtained from both calculations.  In general,
the comparisons are favorable, but distinct differences are observed between
the RGM calculations and some of the polarisation data.  The partial-wave 
decomposition of the experimental data produced by the $R$-matrix analysis
shows that these differences can be attributed to just a few $S$-matrix
elements, for which inadequate tensor-force strength in the $N-N$ interaction
used appears to be responsible.}

\vskip0.4cm
\noindent
PACS numbers: 21.40.+d, 21.60.-n, 25.10.+s

\vskip0.2cm
\noindent
Keywords: $^4$He system, $R$-matrix analysis, RGM calculation, form factors,
phase shifts, reactions

\end{titlepage}

\section
{Introduction}

The $^4{\rm He}$ atomic nucleus is one of the best studied nuclei, both
experimentally and theoretically, as summarized in the recent $ A = 4$
compilation \cite {We1}.  Besides the many textbook
examples of the gross structure, there are subtle points, usually neglected,
which yield large effects. Most of these effects are only qualitatively
and in most cases never quantitatively explained. None of the existing
calculations aims at a complete understanding of the many features of $^4{\rm 
He}$  no
wonder in view of the many different phenomena studied so far \cite {We1}.
With the
recent compilation \cite  {We1} and the comprehensive $R$-matrix analysis 
\cite {Ha1} of a
large amount of scattering data below 10 MeV, a microscopic calculation for
the $^4{\rm He}$ system in this energy range is most needed for.
\vskip 0.2cm
  The difficulties arise already for the total energy of the ground state. In
most cases, parameters of the effective nucleon-nucleon potential are
modified in such a way as to reproduce that energy or the threshold energies,
but seldom both simultaneously.
Calculations using
effective potentials, however, allow to reproduce only some features, while
 failing in
others. Therefore it remains unclear if the failure is due to the deficiencies
of the potential or of the model itself. Only calculations employing realistic
nucleon-nucleon
potentials without any adjustable parameters can help to answer this question.
Such calculations, however, are very complicated and very time consuming,
no matter what type of approach is chosen.
For realistic potentials, calculating the ground state energy is already
a major effort and only three-body forces allow to reproduce that energy within
a few keV in Greens-function Monte-Carlo calculations \cite {Ca1}.
The next open question is the structure of the first excited state, the $0^+$
state just between the $^3{\rm H}-p$ and the $^3{\rm He}-n$ thresholds,
considered frequently to be a breathing mode, which is reasonably
narrow. In calculations this state quite often turns out to be bound
(see e.g. \cite {Ho1}).
On the experimental side this state shows a peculiar behaviour,
indicating a very broad structure in $\alpha - \alpha $ breakup \cite {Reto}.
The $R$-matrix
analysis \cite {Ha1}  indicates the possible origin of this feature.
All the other resonances do not exihibit any narrow structure, but turn out
to have a decay width of the order of MeV \cite {We1,Fi1}, which makes their
location rather arbitrary.
\vskip 0.2cm
  On the other side there is such an amount of scattering data, that
phase shift analyses are possible, especially for proton-triton and
$^3{\rm He}$-neutron scattering.
The $R$-matrix analysis \cite {Ha1}  connects all
the possible reaction channels simultaneously, thus
giving the most complete description and allowing to interpolate to every
desired energy or reaction. Since the number of parameters in this analysis 
is in the order
of one hundred, however, an unrestricted fitting procedure is not possible and
some
physical input is needed, which  will be discussed below.

\vskip 0.2cm
  The aim of this paper is to show that with present-day computers, a rather
unrestricted calculation using realistic nucleon-nucleon potentials,
like the Bonn potential \cite {Mac}, is feasible and yields essentially all
known experimental features. What we consider most surprising is the close
agreement between the $R$-matrix analysis, which has as input only data in the
$^4{\rm He}$  system,
and the refined Resonating Group calculation \cite {Lis}, which
has as input only a realistic nucleon-nucleon potential without any adjustable
parameters.

\vskip 0.2cm

  We organize the paper in the following way: the next chapter
contains a brief review of the Resonating Group Model (RGM) and the model
space used for the calculation, where we want to point out the essential
differences with existing calculations. After that, we review the $R$-matrix
analysis with special emphasis on all restrictions employed in the fitting of 
the data. The next chapter is devoted to a detailed description of the
calculation for the ground state of $^4{\rm He}$.
Then we start with a comparison
of the RGM and $R$-matrix results for the various partial waves. Finally we 
compare selected data with the results of the $R$-matrix analysis and the
RGM-calculation, and give a brief outlook.

\section
{Resonating Group Model and the Model Space for $^4{\rm \bf He}$}

  The Resonating Group Model in its various modifications \cite {Lis,Wil,
Tang} is a suitable
method to calculate the scattering of composite objects. In this case the
main technical problem lies in the evaluation of the many-body matrix
elements.  In order
to facilitate the calculation of the orbital matrix elements, however,
all radial
dependencies have to be of Gaussian form, since only then proper
antisymmetrization for the translationally invariant wave function is feasible.
Therefore also the potential has to be given in terms of Gaussian functions. In
this work we use the $r$-space version of the Bonn-potential \cite {Mac}
expressed in Gaussian functions \cite {Kel}. All further reference to the
Bonn-potential is indeed a reference to this Gaussian version.
\vskip 0.2cm
  Powerful techniques have been developed to achieve the analytic calculation
of the individual matrix elements \cite {Lis, Tang} .
Even with these techniques, calculations are usually far from trivial, hence
most calculations are restricted to single channel problems, neglecting e.g.
effects of the tensor-force, and/or use rather simple effective potentials.
Since the calculation of the orbital matrix elements is the most time consuming
part, symmetries of the orbital wave function are exploited as far as possible
\cite {Lis} . For simple (central or non-central) potentials, single harmonic
oscillator wave functions yield already a good description of the ground state
wave function of the lightest nuclei (see e.g. \cite {Ho1}).
For a realistic $N-N$ potential
this is no longer
the case, because of the strongly repulsive short\-ranged core of
the potential,
which suppresses the wave function for small internucleon distances. 
In many cases even a node occurs in the wave function at small distances. In
terms of Gaussians this means that more than one Gaussian width parameter
is needed to describe even the simplest wave function. Furthermore, the
tensor force plays a crucial role.
\vskip 0.2cm
  Let us explain the situation for the
simplest nucleus, the deuteron: For an effective $N-N$ interaction a single
$S$-wave
binds the deuteron already; for a reasonable radius of the wave function
a linear combination of two Gaussian functions is enough (see \cite {Ho1,Lis}).
For a realistic potential only a superposition of $S$- and $D$-waves yields a
bound deuteron, with at least 3 Gaussians on the $S$-wave and two on the
$D$-wave for the Bonn potential \cite {Mac} (see Appendix A for the detailed
wave function).  For other potentials, e.g. \cite {Eik}, we find the same
behaviour.
\vskip 0.2cm
  In the $^4{\rm He}$ system we have three two-fragment channels, the
triton-proton, the $^3{\rm He}$-neutron, and the deuteron-deuteron channel.
For the latter, the
$D$-wave admixture in the wave function leads to an additional
coupling of internal angular
momenta of 2 to the relative orbital angular momentum. These couplings
lead to very complicated wave functions for the composite system and
increase the computing time neccessary by orders of magnitude. Since we
allow for orbital angular momenta inside the deuteron, we treat the
$^4{\rm He}$
system essentially as a four nucleon system, i.e. we consider it as a
four cluster problem in the framework of the RGM.
\vskip 0.2cm
  Also the triton-proton channel is treated as a four-cluster problem,
because we need internal $D$-waves to bind the triton. To get a reasonable
binding energy, we need $D$-waves on all internal coordinates, thus leading
to a three-cluster description of the triton. Here we need three different
Gaussian width parameters on both internal coordinates. To lower the
binding energy even more, we allow for a second set of width parameters
for all components of the wave function containing at least one $D$-wave.
Some details are given in Appendix A.
Since in $^3{\rm He}$ only a neutron is exchanged
against a proton compared to the triton, the structure of the wave function
is identical, with just modified sets of Gaussian width parameters, given
in Appendix A.

\begin {figure} [h]

\centerline{\epsfxsize=5cm  \epsfbox{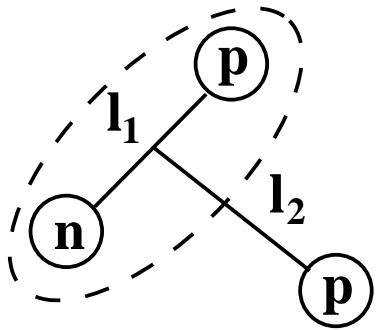}}
\caption{\label{dreh} Angular momentum structure of the $ ^3{\rm He} $ wave
function.}

\end{figure}

Allowing for all possible combinations of $l_1 \leq 2$  and $l_2 \leq 2$
(see fig.\ \ref {dreh}) that can contribute
to $J^\pi=1/2^+$, we get a total of 23 combinations. The total binding energy
and the individal contributions of kinetic and potential energy
 compare favourably well with the results of
Carlson \cite {Ca1} (see table \ref{h3e}). This wave function, however, is too
complicated
to be used in a full-fledged scattering calculation. Therefore we restricted
the triton wave function to the dominant terms. These are all three
spin-isospin combinations
with $l_1=l_2=0$ and furthermore all combinations with the proton-neutron
subsystem
having spin=1 and  pairs $l_1,l_2$ coupled to $l_3$ are $[1,1]2$,  $[2,0]2$,
$[0, 2]2$,
$[2, 2]0$, $[2, 2]1$, and $[2,2]2$. This restricted wave function consists of
only 10 combinations and yields only about 70 keV
less binding energy than the full wave function (see table \ref{h3e}),
 but saves large amounts of computing time.
In test cases we convinced ourselves that this reduction of the wave function
of our physical channels has only minor effects on the $^4{\rm He}$ results.
Also for 
$^3{\rm He}$ we use the similarly restricted wave function.

\begin{table}[t]\caption{\label{h3e} Comparison of triton results with other
calculations and $^3{\rm He}$ results}
\vskip 0.2cm
\centering
\begin{tabular}{c|c|c|c|c|c}
calculation & energy & $ \big < V \big> $ & $ \big < - \frac{\hbar ^2}{2m} 
\Delta \big > $&
$ \sqrt{\big <r^2_i \big >} $ & $ P_{ l_{tot} = 2}$ \\
& [MeV] & [MeV] & [MeV] & [fm]  & [\%] \\
\hline
Carlson(var)& -7.96(03) & -48.2(05) & 40.3(05) & 1.69(02)& 7.0(1) \\
Carlson(Faddeev)& -8.29 & -49.0 & 40.7 & 1.69 &7.0 \\
full wave fuction & -8.015 & -48.875 &40.860 & 1.64 & 6.95\\
reduced w.f. & -7.951 & -48.587 & 40.637 &1.64 & 6.86\\
$^3{\rm He}$ full w.f.& -7.343 & -47.464 & 40.121 & 1.69& 6.89 \\
$^3{\rm He}$ reduced w.f.& -7.258 & -47.124 & 39.866& 1.68 & 6.76 \\
\end{tabular}
\end{table}

\vskip 0.2cm
Together with the deuteron wave function  we find total binding energies and 
threshold energies
which are given in table \ref{thres}. Especially the relative energies compare
favourably well with the experimental numbers.

\begin{table}[bht] \caption{\label{thres} Comparison of experimental and
calculated total binding energies and relative thresholds (in MeV) 
for the reduced
wave functions }
\vskip 0.2cm
\centering
\begin{tabular}{c|c|c|c|c}
channel & \multicolumn{2}{c}{$E_{bin}$} & \multicolumn{2}{c}{$E_{thres}$} \\
 & exp. & cal. & exp. & cal. \\
\hline
$^3{\rm H}-p$ & -8.481& -7.951 & - & - \\
$^3{\rm He} - n $& -7.718& -7.258 & 0.763 & 0.693 \\
$ { d - d} $& -4.448 & -3.809 & 4.033 & 4.142 \\
\end{tabular}
\end{table}

\vskip 0.2cm
All further physical channels are three- or four-body breakup channels, which
cannot  in principle 
be treated within the RGM \cite {Ho2}. To allow for the neccessary amount 
of
flux that goes into these channels, we approximate these breakup channels
by two-fragment channels formed by appropriate combinations of the singlet
deuteron $ \dbar $  with itself or with a deuteron.
The approximate bound state wave function for the singlet deuteron is also
given in Appendix A.
\vskip 0.2cm
This model space, however, is by no means sufficient to find reasonable results
in all partial waves. We have to add so called distortion channels \cite {Ho2},
sometimes called pseudo-inelastic channels \cite {Tang}, to allow for
enough variational freedom in the interaction region.
These distortion channels contain therefore no asymptotic part, so that they
cannot
carry any flux, but have e.g. different symmetries or just increase the
model space such that certain parts of the interaction can yield larger 
contributions. Some details about the
distortion channels will be given during the discussion of the individual
partial waves.
\vskip 0.2cm

Having specified the internal wave functions of our fragments, we indicate the
structure
of the total wave functions; details of the construction of the wave functions
can be found in \cite {Lis,Ho2}. To keep the notation transparent,
we give only the
gross structure of the wave function.
\begin{eqnarray}
\Psi ={ \cal A } \; \left [ \; \sum_{L,S} \left [ \phi_{^3{\rm H}}
\phi_{{\rm p}} \right ]^{S}  \;
\chi ^{S,L}_{^3{\rm H p}} \; 
\left ( {\bf{r}}_{^3{\rm H p}} \; \right ) \;
+ \sum_{L,S} \left [ \phi_{^3{\rm He}} \phi_{{\rm n}} \right ]^{S}  \;
\chi ^{S,L}_{^3{\rm He \, n}} \; \left ( {\bf{r}}_{^3{\rm He \, n}} \right ) \;
\right. \nonumber \\
\left. {} + \sum_{L,S} \left [ \phi_{{\rm d}} \phi_{{\rm d}} \right ]^{S}  \;
\chi ^{S,L}_{{\rm d d}} \; \left ( {\bf{r}}_{{\rm d d}} \right ) \;  
+ \sum \mbox{distortion channels} \; \right ] \;
\end{eqnarray}
Here $ \cal A $ denotes the four particle antisymmetrizer. The relative-motion
wave functions are of the structure
\begin{equation}
\chi ^{L} ( {\bf{r}} ) \; = \left (  b_{L} \cdot f_{L} + a_{L} \cdot
\tilde{g}_{L}
+ \sum_{\nu} c_{\nu L} \cdot E_{\nu L} \right ) \cdot Y_{L} ( \hat {\bf{r}} ),
\end{equation}

where the $f_{L}$ and $\tilde{g}_{L}$ are regular and regularized irregular
Coulomb functions for the appropriate channel and the $E_{\nu L}$ are
Gaussian functions of width $\beta _{\nu}$ times $r^L$, which take care of the
wave function being different in the interaction region from the asymptotic
form. Usually the coefficients $b_{L}$ are determined by the appropriate
boundary condition, i.e. only one element in eq.(1) being unity
and all others zero.
The coefficients $a_{L}$ and $c_{\nu L}$ are linear variational parameters
determined from the Kohn-Hulth\'en variational principle \cite {Kohn}
via
\begin{equation}
\left < \; \delta \Psi \; | H - E | \; \Psi \right > = 0.
\end{equation}
$S$ denotes the channel spin and
$L$ the relative angular momentum.
The $S$-matrix is directly related to the reactance matrix elements $\bf a$,
via the Caley transformation, $ {\bf S} = (1+i {\bf a} ) 
\cdot (1 - i {\bf a})^{-1}$;
for further details of the method see  \cite {Lis,Ho2}.

\section{The $R$-matrix analysis}

  The $R$-matrix formulation \cite {Wig} of multichannel nuclear reaction
theory gives a powerful and convenient energy-dependent framework for
describing the experimental measurements. It parametrizes the unitary
$S$-matrix in terms of (real) reduced-width amplitudes $\gamma_{c \lambda}$
and eigenenergies $E_{\lambda}$, for fixed values of boundary-condition
numbers $B_c$ and channel radii $a_c$. Since these parameters reflect the
nature of the interactions only at small distances ($r_c \le a_c $),
they can be approximately constrained by the symmetry properties of the
strong (nuclear) forces. In this analysis of reactions in the $A = 4$ system,
the approximate charge independence of nuclear forces was used to relate
the parameters in charge-conjugate channels, while allowing simple
corrections for the internal Coulomb effects.
\vskip 0.2cm

The Coulomb-corrected,
charge-independent $R$-matrix analysis of \cite {Ha2} (also used in
\cite {We1}) takes its isospin $T=1$ parameters from an analysis of
 $ p - ^3$He scattering data \cite
{Ha3} that gives a good description of all data at proton energies below
20 MeV. The eigenenergies are, however, shifted by the internal Coulomb
energy difference $\Delta E_{C} = - 0.64$ MeV and the $ p-\rm{^3H}$ and
$ n-\rm{^3He}$
reduced width amplitudes are reduced by the isospin Clebsch-Gordan
coefficient $1/ \sqrt 2$. The isospin $T=0$ parameters are then varied to fit
the experimental data for reactions among the two-fragment channels
$ p-\rm{^3H}$, $ n-\rm{^3He}$, and $ d-\rm{^2H}$, at energies corresponding to
excitations
in $^4{\rm He}$ below 29 MeV. A summary of the channel configuration and data
included for each reaction is given in table \ref {data}.
In this fit, the $T=0$ nucleon-trinucleon reduced-width amplitudes are
constrained by the isospin relation
\begin{equation}
\gamma ^{T=0}_{ n \, ^3{\rm He}} \; = - \gamma ^{T=0}_{ p \, ^3{\rm H}}
\end{equation}
and a small amount of internal Coulomb isospin mixing is introduced by
allowing
\begin{equation}
\gamma ^{T=1}_{{ d d}}   \neq 0
\end{equation}
which is neccessary to reproduce the differences between the two branches
of the $ d-d$ reactions.
Note that this $R$-matrix analysis \cite {Ha2} is not yet completed nor fully
documented. However, this analysis represents the most comprehensive and
detailed attempt to date to give a unified phenomenological description
of the reactions in $^4{\rm He}$. 
\vskip 0.2cm

\begin{table}
\caption{Channel configuration (top) and data summary (bottom) for each
reaction in the $^4$He system $R$-matrix analysis}
\centering
\begin{tabular}{|c|c|c|}      
\hline
Channel  & $l_{\rm max}$ & $a_c$ (fm) \\ \hline
$^3$H$ - p$  &        3  &      4.9   \\ \hline
$^3$He$ - n$ &        3  &      4.9   \\ \hline
$^2$H$ - d$  &        3  &      7.0   \\
\end{tabular}\\
\begin{tabular}{|c|c|c|c|}
\hline
   Reaction      & Energy range (MeV) & \# Observable types & \#
 Data points
\\ \hline
$^3$H$(p,p)^3$H  &    $E_p=0-11$      &          ~3         & 
    1382\\
$^3$H$(p,n)^3$He + inv.& $E_p=0-11$   &          ~5         & 
    ~726\\
$^3$He$(n,n)^3$He&    $E_n=0-10$      &          ~2         & 
    ~126\\
$^2$H$(d,p)^3$H  &    $E_d=0-10$      &          ~6         & 
    1382\\
$^2$H$(d,n)^3$He &    $E_d=0-10$      &          ~6         & 
    ~700\\
$^2$H$(d,d)^2$H  &    $E_d=0-10$      &          ~6         & 
    ~336\\
\hline
                 &     totals:        &          28         &   
  4652\\
\hline
\end{tabular}
\label{data}
\end{table}

Since the level information is derived from a
multilevel
$R$-matrix parametrization, a generalization of the single-level Breit-Wigner
prescription is neccessary. For the convenience of the reader, we repeat here
the essential steps given in \cite {Ha1}.
\vskip 0.2cm
The multilevel generalization of the usual single-level prescription in terms
of $R$-matrix parameters is to find the poles and residues of the
channel-space matrix
\begin{equation}
K_R = G  \left [ \epsilon (E) - E \right ]^{-1} G^T
\end{equation}
in which the elements of G are
\begin{equation}
G_{c \lambda } = P^{1/2}_{c} \gamma _{c \lambda }
\end{equation}
and $\epsilon (E)$ is the level-space matrix of elements
\begin{equation}
\epsilon _{ \lambda ^\prime \lambda} =
E_{\lambda } \delta _{\lambda ^\prime \lambda}
 - \sum _{c} \gamma ^{T}_{\lambda ^\prime c} \left ( S_c - B_c \right )
\gamma _{c \lambda } .
\end{equation}
In eqs.(7) and (8) $S_c$ and $P_c$ are, respectively, the usual
energy-dependent channel shift
and penetrability functions. 
As the notation implies, $K_R$ in eq. (6) is
similar to Heitler's reactance matrix \cite {HH}, except that it is not a true
asymptotic quantity. It therefore gives resonance parameters that depend
on the values of $a_{c}$, so that properties such as level spacings are
functions of the channel radii.
\vskip 0.2cm
Near one of its poles, $K_R$ has the rank one form
\begin{equation}
K_{R} = \frac{1}{2} \frac {  \rho _{R} \rho ^{T}_{R} }{ E_{R} -E}
\end{equation}
in terms of the channel-space residue amplitude $\rho _{R}$
 and pole energy $E_{R}$.
The pole energy is taken to be the ``resonance energy" and the total width is
\begin{equation}
\Gamma = \rho ^{T}_{R} \rho _{R}
= \sum _{c} \rho ^2_{R_{c}} ,
\end{equation}
naturally suggesting that partial widths be defined as

\begin{equation}
\Gamma_c = \rho ^2_{R_{c}}. 
\end{equation}
This prescription gives resonance parameters based on the positions and
residues of {\it apparent} poles of the $S$-matrix, as seen from the real
energy axis of the physical sheet.
\vskip 0.2cm
When the $S$-matrix is continued onto the complex energy surface, near one of
its poles it has the form
\begin{equation}
S = i \frac {  \rho _{0} \rho ^{T}_{0} }{ E_{0} -E}
\end{equation}
where $E_{0} = E_{R} - i \Gamma \slash 2$ is the complex pole energy and
$\rho _{0}$ is the complex residue amplitude. A procedure for obtaining
$E_{0}$ and $\rho _{0}$ from $R$-matrix parameters is given in \cite {Ha4}.
The expectation of the Breit-Wigner approximation is that on any
Riemann sheet, $E_{R}$ will have the value given in eq. (9), $\Gamma$
will be given by eq. (10) and the residue amplitudes $\rho _{0}$ will
differ from $\rho _{R}$
 only by unimportant phase factors. For light systems like the
 $^4{\rm He}$-system
described in terms of $R$-matrix parameters, however, this is often not the
case \cite {Ha4,Ko}.
As explained in \cite {Ha4}, a parameter characterizing the strength of
an $S$-matrix pole,
\begin{equation}
S_{P} =  \frac {  \rho ^{T}_{0} \rho _{0} }{ \Gamma }
\end{equation}
in terms of the magnitude of its residue compared to its displacement from
the real axis can be quite different from unity.

Furthermore, ``shadow" poles \cite {Tay} 
associated with a resonance can have different positions
and residues on different sheets of the Riemann energy surface due to extended
unitarity of the $S$-matrix. We refer to a resonance exhibiting any of these
differences with the Breit-Wigner expectations as a non-Breit-Wigner
resonance.
\vskip 0.2cm
In order to compare the RGM-calculation and the $R$-matrix analysis also in
the level parameters, we adopt the same prescription in both cases.
In the $R$-matrix analysis channel radii $a_{pt}$ = $a_{n \, {\rm ^3He}}$
= 4.9 fm
and $a_{dd}$ = 7.0 fm are used. In the RGM calculation we ``locate" therefore
a resonance at that energy, where the difference between the calculated phase
shift and the ``background phase shift", determined from the appropriate
channel radius, passes through ninety degrees.

\section{The $^4{\rm \bf He}$ Ground State}

In this chapter we will give some information about the ground state of
 $^4{\rm He}$.
For this purpose, we consider different model spaces. The simplest space 
consists
just of the three $^1S_0$ channels  $^3{\rm H}- p$, $^3{\rm He}- n$, 
and ${ d-d}$. In such a
model space we find only half the experimental binding--energy of -28.296 MeV
(see table \ref{Eb}). Adding the $^5D_0$ ${ d-d}$ channel and the
${ \dbar - \dbar} $ channel
to simulate the breakup channels, we gain about 2 MeV (see table \ref{Eb}),
i.e.\ taking all the physical channels into account, the
ground state is only 8 MeV below the lowest threshold.
To improve the binding energy, we have to add distortion channels. First of all
we allow for 30 ${\rm ^3H}- p$ distortion channels and 30 $^3{\rm He}- n$ ones,
which
increases the binding energy by about 8 MeV (see table \ref{Eb}). Adding 82
deuteron-deuteron distortion channels we gain again about 1.5 MeV. Finally we
allowed for additional 62 distortion channels of the structure ${\rm 2N-2N}$,
which brought the energy down to -25.849 MeV, close to the
result of ref. \cite{Ca1}.

\vskip 0.2cm

\begin{table}[t] \caption{\label{Eb} Calculated ground state energies in MeV
for various model spaces}
\vskip0.2cm
\centering
\begin{tabular}{c|c|c|c|c|c|c|c}
\# channels& 3  &5 & 5 + 60 &5+60+82& 5+204& 227 & ref. \cite{Ca1}\\
total energy & -14.033& -15.873 & -24.345 &-25.798 & -25.849 &-25.910 &
-25.86(15)\\
kinetic energy&36.671 &44.834 & 74.767 & 80.642 &80.834 & 80.981 & 81.62(7)\\
Coulomb & 0.611 & 0.641 & 0.772 & 0.795 & 0.796 & 0.797 & 0.74(01) \\
Central & -21.238 &-23.489 & -37.320 & -39.714 & -39.756 & -39.828 & - \\
Tensor & -20.833 & -27.881 & -46.501 & -50.496 & -50.684 & -50.772 & - \\
Spin-orbit & 0.115 & 0.236 & 0.418 & 0.474 & 0.481 & 0.449 & 1.04(01) \\
$p^2$-potential& -9.359 &-10.214& -16.482 &-17.499 & -17.519 & -17.537 &
-17.77(27)\\
\end{tabular}
\end{table}
\vskip 0.2cm

\begin{figure} [t] 

\centerline{\epsfxsize=12cm \epsfysize=12cm  \epsfbox{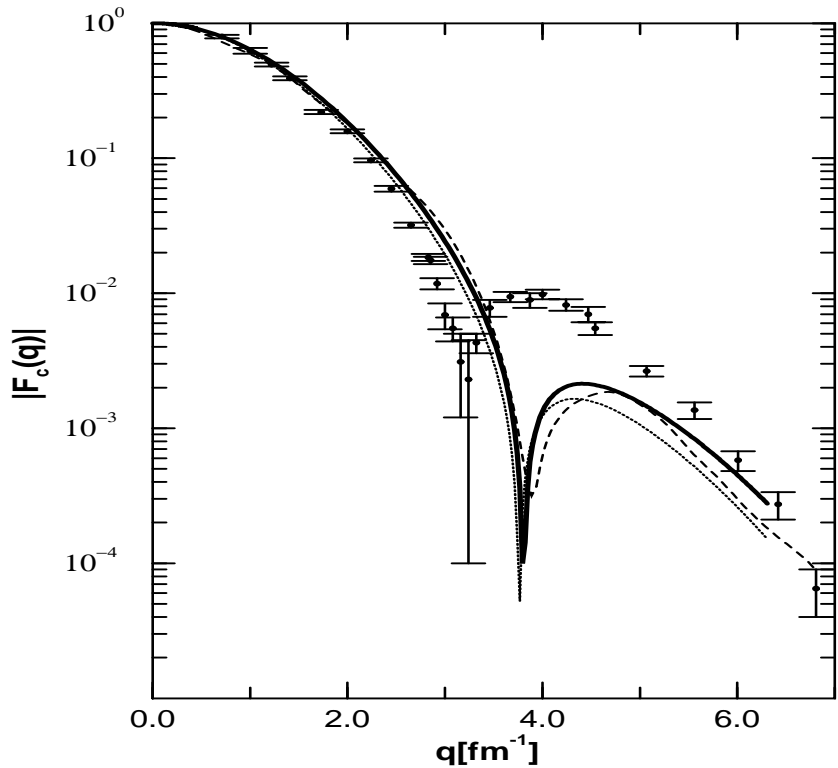}}
\caption{\label{fofa} Point Coulomb form factor for the
$^4{\rm He}$ ground state (full line) together with data \protect \cite{Ar,Fro}
and the results of ref. \protect \cite{Schia} (dashed line). Also shown is
the result for the minimal wave function (dotted line).}
\end{figure}

  Increasing the model space even further improved the binding energy only 
marginally,
but needed much more computing time and caused almost numerical
linear dependencies; therefore we did not pursue it any further.
Since e.g. the triton wave function consists already of 10 different
components, we finally 
diagonalized the Hamiltonian in the full space spanned by all structures
calculated and no longer coupled the various components to physical channels.
The 5 physical 
channels are spanned by 36 different components. We omitted the small width
parameters in these components, which are neccessary to reproduce the
oscillatory behaviour of the scattering wave functions, but correspond to
large distances between the fragments. Hence they lead to minimal interaction
between the fragments and thus yield eigenvalues of the Hamiltonian close to
the threshold energies. We convinced ourselves that a reasonable choice is to
choose the same width parameters as for the distortion channels.
The actual values are given in Appendix B.
Because of numerical linear
dependencies, only 18 out of the 36 components could be retained, yielding a
total of 227 channels. The results are displayed in the next-to-last column of
table \ref{Eb}. The various results are in close agreement with those of ref.\
\cite{Ca1}. The point Coulomb form factor is also quite similar to that of
ref. \cite{Schia} found for the Argonne potential (see fig.\ \ref{fofa}).
Therefore we
believe that the inclusion of meson-exchange-currents into our calculation
would also reproduce these data.

\vskip 0.2cm 
Our calculation with 227 channels corresponds to a
calculation in a certain configuration space with only bound states. Hence, we
consider
the two lowest eigenvalues of the Hamiltonian in this space as variational
approximations to the ground and first excited  $0^+$ states. 
In table \ref{qb} we give the binding energy, the probability of total nucleon
spin being equal to two, that for it being equal to one, the charge radius
and the root-mean-square radius for the ground state and the first excited
state. The probability of $ S=2$ is often called $D$-state probability, that of
$ S=1 $ sometimes $P$-state probability. Since,
however, various orbital angular momenta of two can be coupled to a total
angular momentum of zero, which is non-orthogonal to a pure $S$-state,
we consider the above definition the better one.
\begin{table}[hbt] \caption{ \label{qb} Properties of the first two $0^+$states
in various model spaces}
\vskip 0.2cm
\centering
\begin{tabular}{c|c|c|c|c|c}
         & full space & coef$ >$0.01 & coef$ > $0.015 & minimal & ref.
	 \cite{Ca1}\\ \hline
$ \# {\rm channels} $ & 227 & 138 & 58 & 20 & \\
$E_{tot} (MeV)$ & -25.910 &-25.276 & -24.557 & -23.203& -25.86(15)\\
$ P_{S=2} (\%)$& 10.2 & 10.5 & 10.1 & 9.1 & 11.3(1)\\
$ P_{S=1} (\%)$ & 0.19 & 0.18& 0.13 & 0.08& - \\
$r_{ch} (fm)$& 1.47 & 1.49& 1.51 & 1.55& - \\
$r_{rms} (fm)$& 1.47 & 1.49 & 1.51& 1.54& 1.50(1)\\
\hline
$E_{0^+} (MeV) $ & -6.417 & -6.205 & -5.797 & -5.245 & - \\
$ P_{S=2} (\%) $ &6.86 & 6.79 & 6.59 & 6.34& - \\
$ P_{S=1} (\%) $ & 0.06 & 0.06 & 0.03 & 0.02& - \\
$ r_{ch} (fm) $ & 3.09 & 3.10 &3.08 & 3.04& - \\
$r_{rms} (fm) $ & 3.02 & 3.02 & 2.99 & 2.97 & - \\ \hline
\end{tabular}
\end{table}
\vskip 0.2cm
In the first column of table \ref{qb}, we give the results of the full
calculation.  All results for the ground state 
agree reasonably well with those of ref. \cite{Ca1}, although the
probability of finding spin two is a bit smaller, which might be due to
slightly different parameters for the potential used.
The $ S=1$ configurations contribute about one MeV to the binding energy,
despite their marginal admixture. Due to the different width parameters used
for the triton and $^3{\rm He}$ fragments, the charge and mass radii of
$^4{\rm He}$ can in principle be different. The differences, however, are
too small to show up in table \ref{qb}.
The energy of the first excited state is above the $^3{\rm He}- n$ threshold.
Allowing for all width parameters used for the physical channels, it would
occur a few keV above the lowest threshold, the $^3{\rm H}-p$ one.
Thus the first excited state would no longer be a reasonable approximation
to the $0^+$ resonance, but rather
of the lowest threshold. We convinced ourselves that
the model space chosen yields a ground state energy just 60 keV less bound than
in the space containing all width parameters. This is about the same amount
gained by uncoupling the physical channels. Considering, however, how much
the model space is increased, the gain is just a third of the former.

\begin{figure} [t] 
\centerline{\epsfxsize=12cm \epsfysize=12cm  \epsfbox{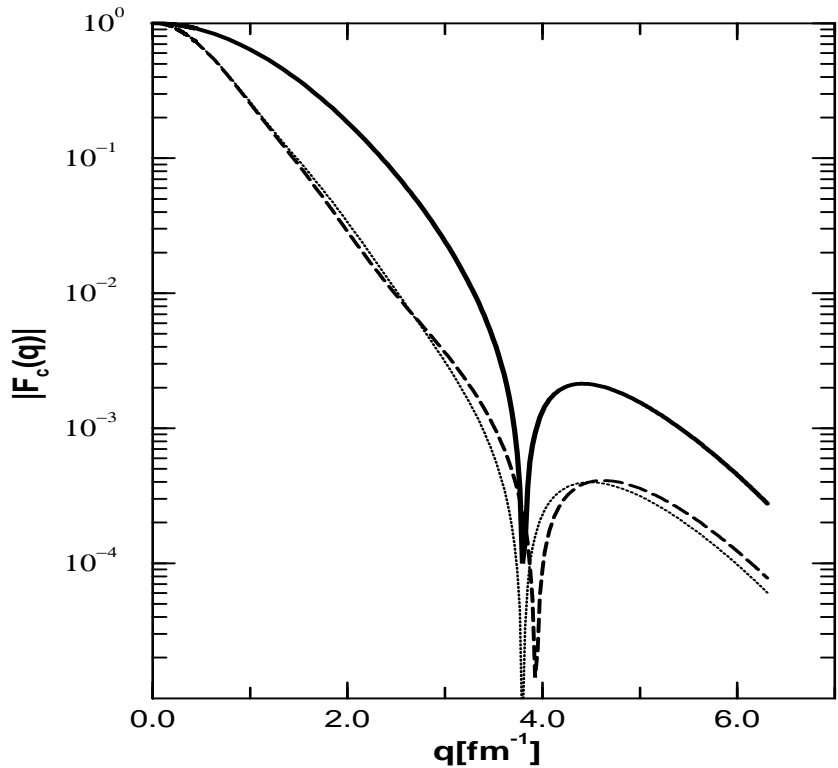}}

\caption{\label{fofa1} Point Coulomb form factor for the
first excited state (dashed line) compared to the ground state form factor
(full line).  Also shown is
the result for the minimal wave function (dotted line).}
\end{figure}
\vskip 0.2cm

If we reduce, however, the number of width parameter to only four, then
we lose only 25 keV for the ground state, but three MeV for the excited state.
Therefore we consider the model space chosen as optimal.
The relatively high energy found for this excited state might explain the
difficulties of shell model calculations \cite{Ceu} to reproduce the position
given in the compilation \cite{We1}.
\vskip 0.2cm

As one might have expected,
the radius of the first excited state is much larger than that
of the ground state.
The probabilities for spin unequal to zero, however, are much lower than for
the ground state, in contrast to the results found in ref. \cite{Ceu}, 
shading doubts on the interpretation of this state
as a breathing mode \cite{Ceu}. Also the form factor for the first excited
state is quite different from that of the ground state (see fig.\ \ref{fofa1}).
The minimum is almost at the same position as for the ground state, but for
most momentum transfers the form factor of the first excited state is about
an order of magnitude below the ground-state one.

\begin{figure} [t] 
\centerline{\epsfxsize=12cm \epsfysize=12cm  \epsfbox{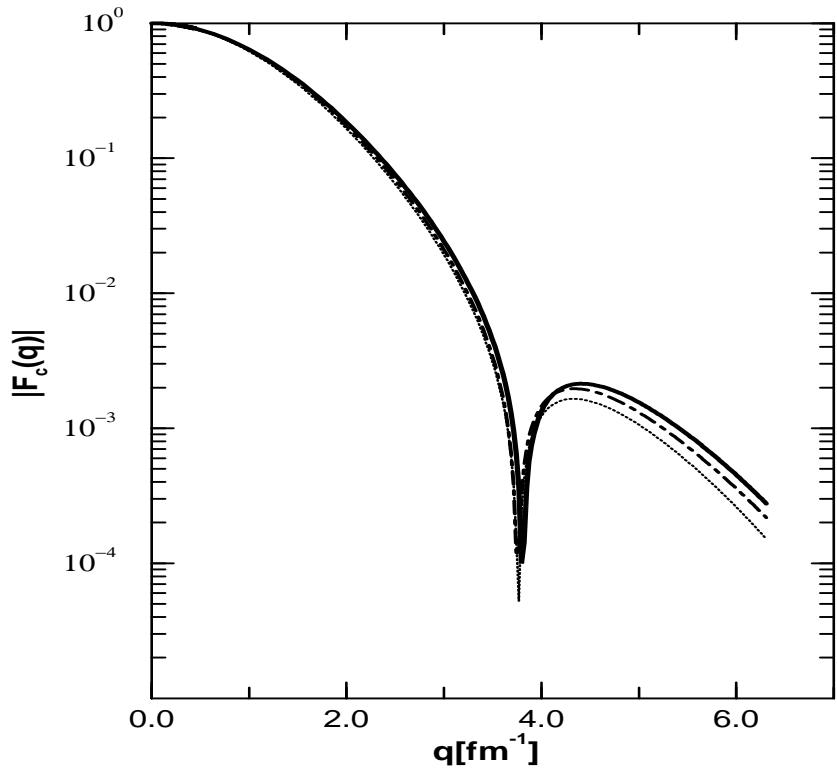}}
\caption{\label{fofav} Comparison of the 
point Coulomb form factor of the ground state for the full wave function
(full line), the 58 channel wave function (dashed line) and the minimal
wave function (dotted line).}
\end{figure}

Since a wave function consisting of more than 200 components, each containing
five different width parameters, is not instructive, we tried to find a much
simpler one by reducing the number of configurations. The most obvious idea
of omitting all $S=1$ components, because of their tiny percentage, was not
successful, as we lost well over one MeV in binding energy. Therefore we
omitted all configurations which had (non-orthogonal) expansion coefficients
below 0.01.
Thus reducing the number of channels by roughly 40 percent, we lost only a
moderate amount of binding energy, but all the other results remained almost
unchanged (see second column in table \ref{qb}). The change in the form
factors is too small to be displayed in fig.\ \ref{fofa}. 
Omitting repeatedly all configurations with
coefficients not exceeding 0.015 we could again cut the number of channels
by more than half. The binding energies are now already reduced
appreciably, but the other results hardly change (see third column in table
 \ref{qb}); also the form factors are almost unchanged (see fig.  \ref{fofav}).

\vskip 0.2cm
  This wave function, however, is still too complicated to be used in, say,
radiative capture calculations. Omitting the last remaining
$S=1$ channels, which had expansion coefficients barely larger than those of
the deleted states, we lost again more than an MeV. Therefore we changed the
strategy of channels to be deleted: We deleted every channel if the loss
of binding energy was below 70 keV. Employing this procedure we ended up
with just 20 channels. We call this the minimal wave function, because
deleting any additional channel reduces the ground state binding energy
by more than one hundred keV and, more importantly, the energy of the first
excited state by more than 350 keV. The results for this wave function are
displayed in table  \ref{qb} and figs.\ \ref{fofa}, \ref{fofa1} and
\ref{fofav}.  Here the loss in binding energy is large,
but all other calculated quantities are still reasonable. Therefore we
consider this wave function useful as a starting point for more detailed
studies, like ${ d-d}$ radiative capture \cite{Ed} .
Its structure is given in appendix B.

\section{Partial Wave Analysis}

  In the following we will present the results of the resonating group
calculation and the $R$-matrix analysis for all relevant partial waves. Because
of the large amount of data we restrict the presentation mainly to the
elastic scattering phase shifts. Only in cases where these seem to disagree,
we present some additional information, like Argand plots. The
$R$-matrix results are given for an energy up to 7.5 MeV in the center-of-mass
of the triton-proton
channel and correspondingly in the other channels. This energy is well
within the range of data taken into account, and therefore end-of-data effects
should be negligible. To give an impression how the calculated phase shifts
vary with energy, we display the RGM results up to an energy of 12.5 MeV.
The scattering calculation, however, is stable to the highest energy calculated
till now, of 80 MeV. We consider as ``physical channels'' in the RGM
calculation the $ { t - p}$ and $^3{\rm He }- n$ channels and all
[2+2] fragmentations, like
${ d - d}$, ${ d - \dbar}$ and ${ \dbar - \dbar}$, which can be coupled to the
quantum numbers considered.

\begin{figure} [t]
\centerline{\epsfxsize=10cm  \epsfbox{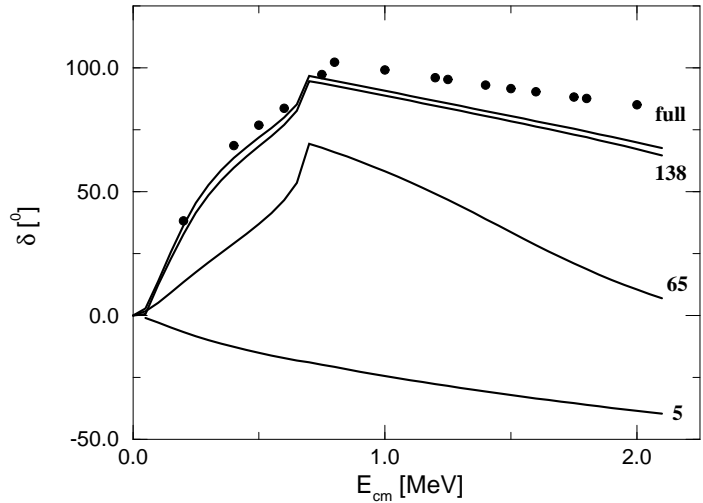}}

\caption{\label{np1} Comparison of elastic $ ^1S_0 $ ${t - p} $ phase shifts
extracted from $R$-matrix analysis (full dots) and calculated for various 
model spaces within the RGM (full lines) near the $ ^3{\rm He }- n $ threshold.
The numbers besides the lines denote the dimension of the model space. Note
that the calculated threshold is a bit lower than the experimental value,
(cf. table \protect {\ref{thres}}.)}
\end{figure}

\vskip 0.2cm
  Let us start with the partial waves of positive parity. The most elaborate
one to calculate is for the $0^+ $-channels, because in this partial wave there
is the bound ground state and the low-lying first excited state. In fig.\ 
\ref{np1} we compare the $ {t -p }$ phase shifts extracted from the $R$-matrix
analysis with RGM-calculations for various model spaces in the neighbourhood
of the $^3{\rm He} - n $ threshold. The $R$-matrix results show a steep rise
till the $ ^3{\rm He }- n $ threshold exceeding 100 degrees and a gradual
decrease above the threshold.

\vskip 0.2cm
  For the RGM, we give results for model spaces similar to the ones used for
the bound state (see table \ref {Eb}). For the simplest model space of 5
physical channels, as described in the previous section, we find a repulsive
triton-proton phase shift. The $ ^3{\rm He} -n $ threshold is hardly
detectable.  Just around 8 MeV there is a sign of the first excited state
(see table \ref{levs}).

\begin{table} [t]
\renewcommand{\arraystretch}{0.75}
\caption{\label{levs} Resonances extracted from the $R$-matrix approach and the
RGM calculations. In the first column the $J^{\pi}$ value is given, then the
dominant isospin. In the third column the dominant structure is indicated, if
it is nontrivial, then the energy relative to the ${ t-p}$ threshold as given
in \protect \cite{We1}. In the next two columns we give the energies of the
latest RGM calculation (usually the one with the largest model space) and
one, which allowed only for the few physical channels as described in the text.
All energies are given in MeV.}

 \vskip 0.4cm

\centering
 \begin{tabular}{c|c|c|c|c|c}
 $J^{\pi}$ & T & struc & $R$-matrix & RGM full & RGM phys \\ \hline
 $0^+$& 0 & & 0.4 & 0.5 & 7.7\\
 $0^-$& 0 & & 1.2 & 1.2 & 2.2 \\
 $2^-$& 0 & & 2.0 & 2.9 & 4.2 \\
 $2^-$& 1 & & 3.5 & 4.5 & 5.7 \\
 $1^-$& 1 &$^3P_1$ & 3.8 & 4.6 & 5.6 \\
 $1^-$& 0 &$^3P_1$ & 4.4 & 5.1 & 6.0 \\
 $0^-$& 1 & & 5.5 & 5.8 & 7.0 \\
 $1^-$& 1 &$^1P_1$ & 6.1 & 6.7 & 7.5 \\
 $2^+$& 0 & & 7.6 & 7.3 &  \\
 $1^+$& 0 & & 8.5 & 11.0 &  \\
 $1^-$& 0 &$d-d$ & 8.6 & 12.7 & 13.5 \\
 $2^-$& 0 &$d-d$ & 8.6 & 11.5 & 12.2 \\
 $0^-$& 0 &$d-d$ & 8.8 & 10.9 & 11.4 \\
 $2^+$& 0 &$d-d$ & 8.9 & 10.7 &  \\
 $2^+$& 0 &$d-d$ & 10.1 & 11.0 &  \\
 $0^+$& 0 &$^5D_0$ & - & 10.9 & 11.0 \\
 $0^+$& 0 &$d-d$ & - & 10.3 & - \\
 $1^+$& 0 &$d-d$ & - & 11.0 &  \\
 \end{tabular}
\end {table}

Adding the 60 distortion channels of [3+1]-fragmentations yields qualitatively
the $R$-matrix result: a steep rise up to the $ ^3{\rm He} - n $ threshold,
 which is
a bit lower than in reality (compare table \ref{thres}), and then a gradual
falloff.
The possibility of a Wigner cusp at the threshold cannot be
decided from the calculation. From the steep rise one could conclude 
the existence of a nearby $ 0^+ $resonance above the threshold. The full
calculation, which includes 204 distortion channels of various kind
 (see previous
section), reproduces the $R$-matrix results below the $ ^3{\rm He} - n $ 
threshold
almost perfectly, reaches 100 degrees at threshold (which is a bit too low)
and then falls off gradually.
From this calculation, we would position the
first excited state in agreement with the $R$-matrix just below the threshold
(see table \ref{levs}).
In the previous chapter we showed that the bound-state calculation
for a model space containing the same structures, but fewer width parameters
on the relative coordinate of the physical channels, puts the state well
above the $ ^3{\rm He} - n $ threshold. This means that the state is strongly
shifted due to the coupling to the open channels, a fact also found in the 
$R$-matrix analysis (see ref. \cite {We1}). As was mentioned in ref.
\cite {We1}, the $S$-matrix pole corresponding to the first excited $0^+$
state in the $R$-matrix analysis is quite broad, and appears well above the
$ ^3{\rm He} - n $ threshold, at about 3MeV relative to the $ t- p$ mass. In 
these respects, it differs considerably from the resonance parameters
determined by the usual prescription (the multi-level generalisation of which
we have used), and from the apparent position and width of the resonance in the
data. However, this is clearly the case of a non-Breit-Wigner resonance, as
discussed in section 3, that appears in the $^4 {\rm He}$ data primarily as a
narrow threshold effect (near the $^3{\rm He} -n $ threshold), but is in fact a
relatively broad structure located higher in excitation energy, as it appears
in $ \alpha - \alpha $ scattering \cite {Reto} and in most calculations,
including the present one.

\vskip 0.2cm

\begin{figure} [t]
\centerline{\epsfxsize=10cm  \epsfbox{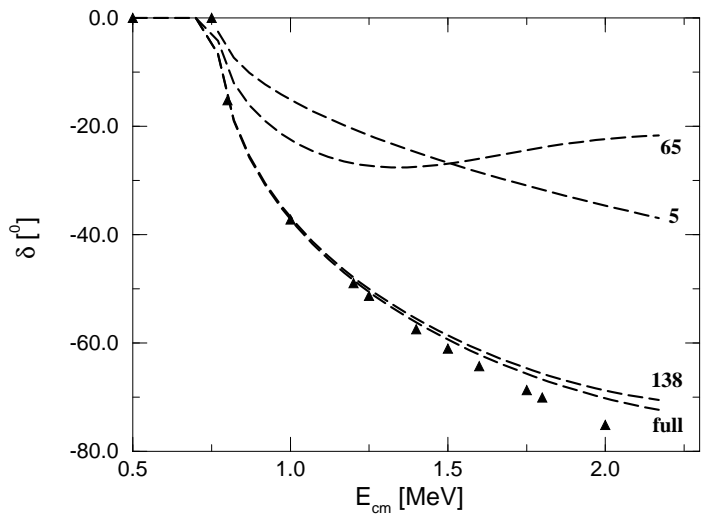}}

\caption{\label{np2} Comparison of elastic $ ^1S_0 $ $^3{\rm He} -n$ phase
shifts extracted from $R$-matrix analysis (full triangles) and calculated
for various model spaces within the RGM (dashed lines) above the
$ ^3{\rm He }- n $ threshold.
The numbers besides the lines denote the dimension of the model space used.}
\end{figure}

 This full calculation seems to be
the optimal one. Near the deuteron-deuteron threshold, however,
numerical instabilities develop, because of linear dependencies in the
[2+2]-fragmentations.
In order to get numerical stability we had to remove
most of the [2+2] distortion channels. Keeping 133 distortion channels
yielded numerically stable results for the whole energy range calculated.
As can be seen in fig.\ \ref{np1}, the deviations from the full calculation
are only minor ones.

\vskip 0.2cm

  In fig.\ \ref{np2} we present the elastic $ ^3{\rm He} - n $ scattering
phase shifts again from the $R$-matrix analysis and the various model spaces
in the RGM. Starting at the threshold the $R$-matrix phase shifts are
negative and fall to about -80 degrees at 2 MeV. Note that in the following 
all energies are
center-of-mass energies above the triton-proton threshold.

\begin{figure} [t]
\centerline{\epsfxsize=10cm \epsfbox{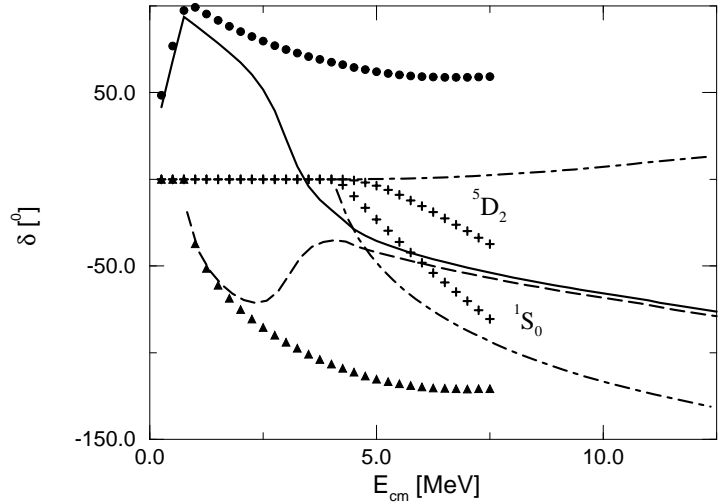}}

\caption{\label{np3} Elastic $0^+$ phase shifts for all physical two-fragment
channels. The data from the $R$-matrix are denoted by full dots ($ {t - p} $
channels), full triangles ($ ^3{\rm He} - n $ channels), and crosses
($ { d - d} $
channels). The calculated phase shifts are denoted by full lines, dashed
lines, and dashed-dotted lines, respectively. For fragmentations with more
than one channel, the quantum numbers are indicated.}
\end{figure}

For the physical channels alone, the RGM yields also negative phase
shifts, but far from the $R$-matrix ones.  Also the calculation including the
[3+1] distortion channels is still far from the $R$-matrix data. The full
calculation, however, reproduces the data again almost perfectly and the
reduction to 138 channels yields only minor modifications at the higher
energies (see fig.\ \ref{np2}).

\vskip 0.2cm
  In fig.\ \ref {np3} we display the phase shifts of all two-fragment
channels over the whole energy range. The beautiful agreement at low energies 
between the $R$-matrix and RGM results obviously does not carry on to higher
energies. (The RGM calculation is the 138 channel one.) 

\vskip 0.2cm
  Let us now consider the ${ d - d}$ phase shifts. The 
calculated  $ ^1S_0$
phases agree reasonably well with those from the $R$-matrix. The $^5D_0$
phases are, however, qualitatively different: the $R$-matrix phases are
negative and reach more than -30 degrees at 7.5 MeV, whereas the calculated
phase shifts are tiny and positive. We will discuss this point in the final
section.

\vskip 0.2cm

  The calculated [3+1] phase shifts start to deviate from the $R$-matrix
results around 2 MeV. Whereas the $R$-matrix phase shifts for the two channels 
differ by about 180 degrees, a difference which does not modify any
experimental datum, the RGM phases for the two channels converge to almost
the same negative value, however, different from the $R$-matrix result (see
fig.\ \ref {np3}).

\begin{figure} [t]
\centerline{\epsfxsize=10cm  \epsfbox{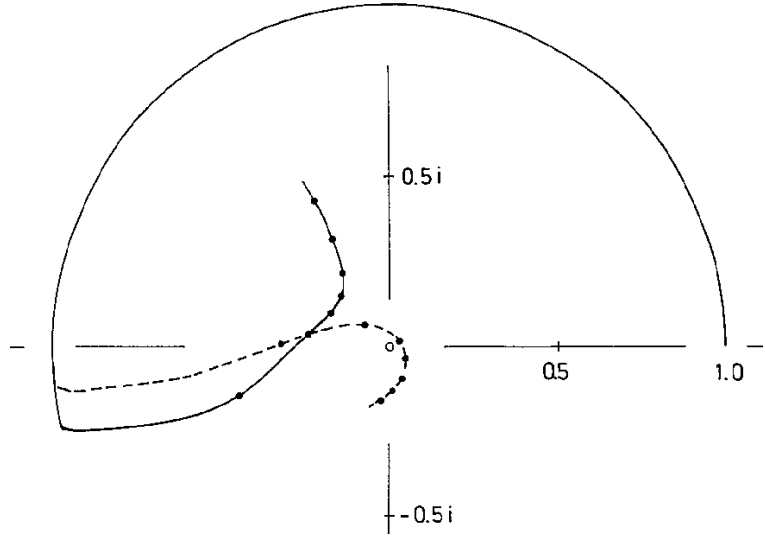}}
\caption{\label{np4} Argand plot of the $ ^1S_0$ $ {t - p} $ $S$-matrix
element. The full line represents the $S$-matrix deduced from the
$R$-matrix in the complex plane, the broken line
the one calculated within the RGM. The dots on the curves denote
steps of one MeV in the center-of-mass energy.}
\end{figure}

The Argand plot of the $ {t - p} $ $S$-matrix elements (see
fig.\ \ref{np4})
hints at the origin of this difference. The diagonal $S$-matrix
elements become (very) small in both approaches, due to the underlying
$ 0^+$, $T=0 $ resonance, discussed above. Since this resonance has almost
good isospin $T=0$, the coupling $S$-matrix element between the triton-proton
and the $ ^3{\rm He }- n $ channel reaches almost the unitary limit of unity.
Even at 7.5 MeV the coupling matrix element exceeds the diagonal ones
appreciably in both approaches. Since the diagonal $S$-matrix elements are
so small, elastic scattering will not allow to determine them more precisely
(see discussion below).
\vskip 0.2cm

From an
analysis of the calculated eigenphase shifts together with the background
phases for channel radii as given in ref \cite {We1}, we deduce two additional
resonances of almost pure ${ d-d}$ structure between 10 and 11 MeV. These 
resonances do not occur in the recent compilation \cite {We1}. The one of
$^5D_0$ structure appears already in the model space of only physical channels
(see table \ref{levs}).
These high lying resonances of pure ${ d-d}$ structure
will occur in more partial waves and we will discuss them at the end of this 
section.

\begin{figure} [t]
\centerline{\epsfxsize=10cm  \epsfbox{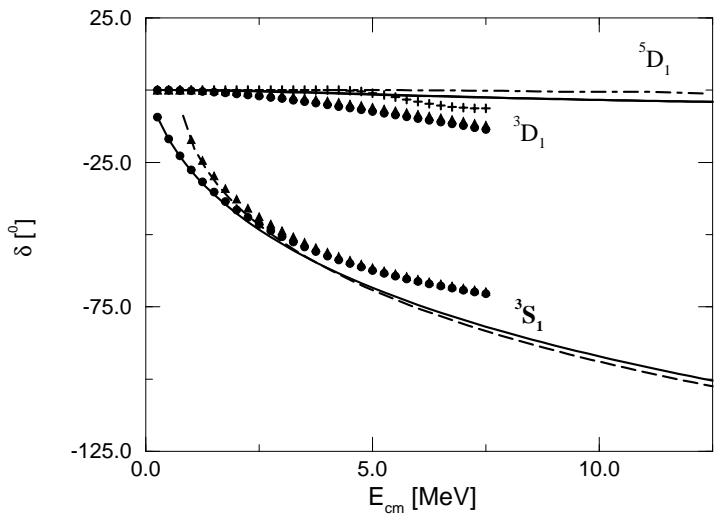}}

\caption{\label{ep1} Same as fig.\ \protect \ref{np3},
but for the  $1^+$ phase shifts.}
\end{figure}
\vskip 0.2cm

  Let us now come to the next partial wave, the $1^+$ one.
In this partial wave the
$R$-matrix analysis \cite {We1}
finds only one broad resonance about 9 MeV above the
$ {t - p}$ threshold. Therefore we restricted the RGM calculation to the
physical channels only, the $ ^3S_1 $ and $ ^3D_1 $ channels for the
[3+1] fragmentations and $ ^5D_1 $ for the deuteron-deuteron channel. In fig.\ 
\ref {ep1} we compare the elastic phase shifts from the $R$-matrix and the
RGM for all five channels. All $R$-matrix phases are negative, with the
$S$-phase shifts reaching -70 degrees, whereas all $D$-phases do not even
reach -15
degrees. The calculated phase shifts are negative too. The $S$-phases agree
nicely with the $R$-matrix ones and the $D$-phases are even smaller.
From an
analysis of the calculated eigenphase shifts together with the background
phases for radii as given in ref \cite {We1}, we again try to deduce
resonances. We find in the energy range considered solely
a single broad $ T = 0 $
resonance around 11 MeV (see table \ref{levs}).
This is appreciably higher than in the $R$-matrix
analysis \cite {We1}, but the resonance has the same structure.

\begin{figure} [t]
\centerline{\epsfxsize=10cm  \epsfbox{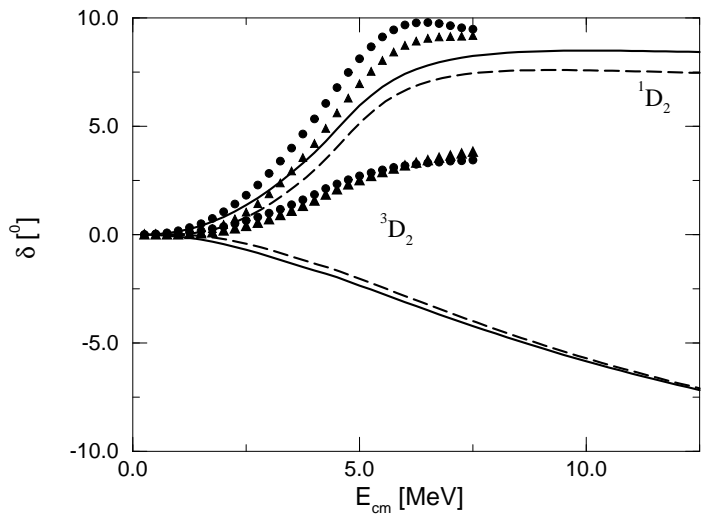}}

\caption{\label{zp1} Comparison of $2^+$ $ { t - p} $ and $ ^3{\rm He }- n$
phase shifts.}
\end{figure}

Since we did not allow for distortion channels, which lead to a stronger
coupling of channels in the interaction region, we anticipate that the energy
of this resonance would be reduced when distortion channels are included,
so that the resonance energies agree more closely.

\vskip 0.2 cm

 The $2^+$-partial wave contains the most coupled channels. In the $R$-matrix
analysis
the $^1D_2$ and $^3D_2$ [3+1] fragmentation channels are taken into account
together with the $^5S_2$, $^1D_2$, and $^5D_2$ deuteron-deuteron channels.
The $^5G_2$ ${ d - d} $ channel is neglected. In the RGM calculation, we again
restrict ourselves to the physical channels and allow
for the same 7 channels. A test calculation \cite {Ed} including the $^5G_2$
channel, but partially simpler internal functions, yields only minor
modifications
and is therefore not discussed here. In fig.\ \ref {zp1} we compare the
$R$-matrix [3+1] phase shifts with those calculated from the RGM. All
$R$-matrix
phases are positive, the $^1D_2$ reaching almost 10 degrees, with the
$ {t - p} $ phases a bit larger than the $ ^3{\rm He} - n $ ones. The $ ^3D_2 $
phases are quite small, barely reaching 3 degrees. The $^1D_2$ RGM phase
shifts agree nicely with the $R$-matrix ones, even so they do not reach
quite as high. The $^3D_2$ phases, however, have essentially the opposite
sign. It should be noted, however, that a previous $R$-matrix analysis
using a somewhat smaller data basis also yielded negative $^3D_2$ phase
shifts. The origin of this sign change is not yet known.

\vskip 0.2cm

\begin{figure} [t]
\centerline{\epsfxsize=10cm  \epsfbox{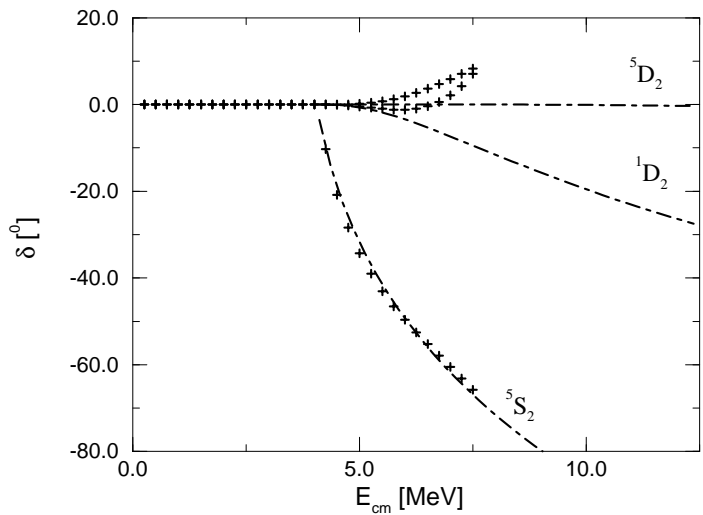}}

\caption{\label{zp2} Comparison of $2^+$ $ { d - d} $
phase shifts.}
\end{figure}

  In fig.\ \ref{zp2} we present the $ { d - d }$ phase shifts. The $ ^5S_2$
phases fall off strongly and agree perfectly for the two approaches. All 
$D$-wave phase shifts are small. The $R$-matrix analysis finds positive values
at the higher energies, whereas the RGM finds negative values.

\vskip 0.2cm

  In the recent compilation \cite {We1} three $ 2^+ $ $ T=0$ levels are
given, all essentially of rather pure ${ d-d}$ structure, at 7.6, 8.9,
and 10.1 MeV above the  ${t -p}$ threshold, having primarily $^5S_2$, $^1D_2$,
and $^5D_2$ components. In the RGM calculation, we
cannot easily find the complex energy poles of the $S$-matrix, so that  
the partial widths of the various channels can only be given in case of
a Breit-Wigner resonance. We can, however, always diagonalize
the $S$-matrix to get the eigenphases and the eigenvectors at all real
energies.  From these eigenvectors we can determine the isospin value
and even its purity of
the level under consideration and its structure.
In the energy range considered, we find also three $T=0$ resonances,
having predominantly ${ d-d}$ structure, at 7.3, 10.7, and 11.0 MeV. The 
structure is the same as found in the $R$-matrix analysis, but the state
at 10.7 MeV has also components of $^1D_2$ ${t-p}$ and $^3{\rm He}-n$ of about
30 percent. Since the width of all these resonances is some MeV \cite {We1},
the agreement is quite good (see table \ref{levs}).

\vskip 0.2cm

  For the $3^+$ and $4^+$ partial waves, the $R$-matrix analysis finds small
negative values for all
$D$-wave phase shifts for all fragmentations. Till now there is no
full-fledged RGM calculation for these partial waves. In order to get an
idea what might come out of such a calculation, we used all the matrix
elements calculated so far that could be coupled to $3^+$ or $4^+$.
For the [3+1] fragmentations, the resulting internal wave functions are
quite good, which can be deduced from the resulting threshold energy, where
we loose some hundred keV only. The deuteron-deuteron channels, however,
are almost unbound relative to the four-nucleon threshold
in the $3^+$ case and unbound by more than 3 MeV for
the $4^+$ channel. This is also reflected in the calculated phase shifts.
The $^3D_3$ [3+1] phase shifts are negative and close to their $^3D_1$ and
$^3D_2$ counterparts. The $^5D$ ${ d-d}$ phases are tiny
and even positive in the $4^+$ channel.

\vskip 0.2cm

  Before presenting the negative parity partial waves, let us briefly
summarize the results for all $D$-waves. For the $^3D_J$ [3+1] phase shifts,
the $R$-matrix yields negative values for $J=1$ and $J=3$, but positive ones
for $J=2$. Such a behaviour can be due to a strong tensor force. The RGM
calculation yields negative values in all cases with barely any $J$-splitting.
Also the splitting of the $R$-matrix 
$^5D_J$ $ { d - d}$ phase shifts follow essentially
the pattern of a strong tensor force. The corresponding  RGM ones are tiny
and show almost no splitting.

\vskip 0.2 cm

\begin{figure} [t]
\centerline{\epsfxsize=10cm  \epsfbox{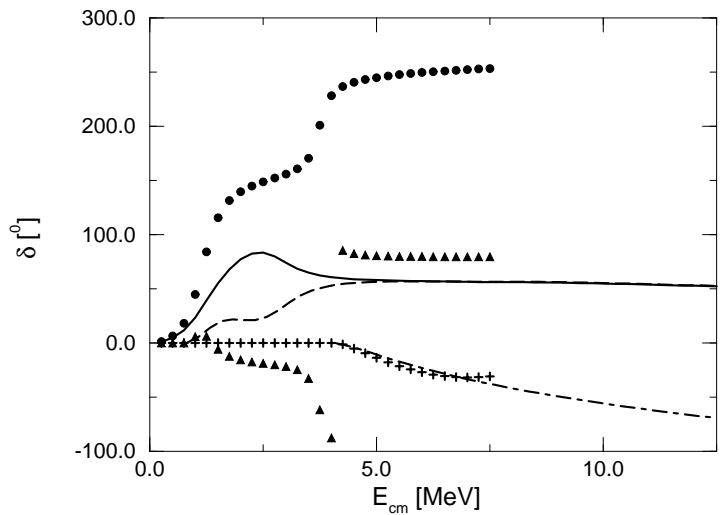}}

\caption{\label{nm1} Same as fig.\ \protect \ref{np3},
but for the  $0^-$ phase shifts.}
\end{figure}

\begin{figure} [t]
\centerline{\epsfxsize=10cm  \epsfbox{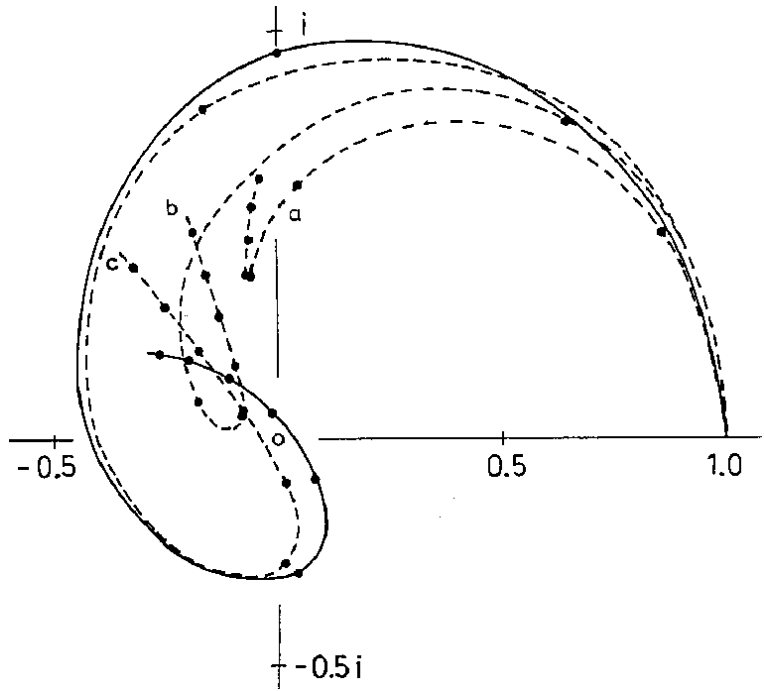}}
\caption{\label{nm2} Argand plot of the $^3P_0$ elastic ${t -p}$ $S$-matrix
elements extracted from the $R$-matrix (full) and calculated by using the
physical channels only (dashed line a), adding 6 ${ d-d}$ distortion channels
(dashed line b) and adding a total of 46 distortion channels (dashed line c).
The origin is marked as a small circle in the middle of the figure.}
\end{figure}

\begin{figure} [t]
\centerline{\epsfxsize=10cm  \epsfbox{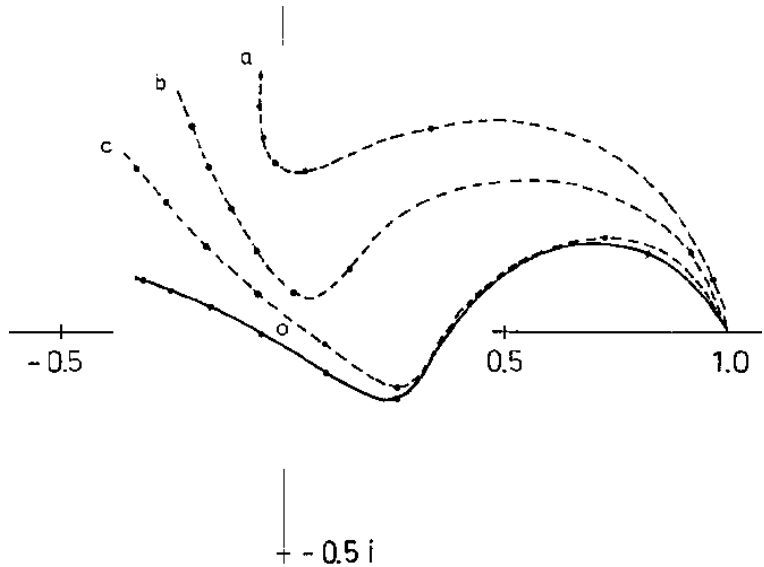}}
\caption{\label{nm3} Argand plot of the $^3P_0$ elastic $^3{\rm He}-n$
$S$-matrix
elements extracted from the $R$-matrix (full) and calculated by using the
physical channels only (dashed line a), adding 6 ${ d-d}$ distortion channels
(dashed line b) and adding a total of 46 distortion channels (dashed line c).}
\end{figure}

  Let us now come to the negative-parity partial waves. In the compilation
\cite {We1} three $0^-$ resonances are given. The $R$-matrix analysis takes the
$^3P_0$ for the three physical channels into account.
In fig.\ \ref{nm1} the $R$-matrix phases are compared to the RGM ones. From the
steep rise of the $ {t - p} $ $R$-matrix data around 1.0 and 4.0 MeV, one can
easily conclude the existence of two resonances. The third one, which is
of ${ d - d}$ structure \cite {We1} is not apparent. Except where the [3+1]
phases vary rapidly with energy, the $ {t - p} $ and $^3{\rm He} - n $ phases
differ essentially by multiples of 180 degrees. Above 4.5 MeV we have added
180 degrees to the $ ^3{\rm He} - n$ phase shifts 
in oder to facilitate the comparison with
the RGM results.
\vskip 0.2cm
  The RGM calculation takes also the physical channels into account and a
few distortion channels of ${ d- d}$ structure. As for the $0^+$ channels,
the [3+1] results look obviously different for the lower
energies. At the higher energies they do not quite reach the $R$-matrix phases.
The $ { d - d}$ phase shifts agree nicely. Similar to the $0^+$ partial waves,
we anticipate that the resonances with their rather pure isospin are the
origin of the differences. As can be seen from the Argand plots in figs.\ 
\ref {nm2} and \ref {nm3}, 
this is indeed the case. The model space used here corresponds to
curve $b$ in figs.\ \ref {nm2} and \ref {nm3}. 
Even though the agreement is not perfect,
all the qualitative features agree. Since the $R$-matrix and RGM curves of the
$S$-matrix pass on different sides of the origin, one of the phases is
 increasing and
the other decreasing. From the eigenphase shifts we can deduce also
three resonances, two of [3+1] structure; a $ T = 0 $ one at 1.6 MeV which
has a rather large $T=1$ admixture
and a rather pure $ T = 1 $ one at 6.6 MeV, and one of $ { d - d} $
structure at 11.1 MeV. These findings
agree again nicely with those of ref. \cite {We1} (cf. table \ref{levs}).

\vskip 0.2cm

  In order to demonstrate the effect of changing the model space, we display
in figs.\ \ref {nm2} and \ref {nm3} the $S$-matrix elements for elastic
triton-proton and $ ^3{\rm He} - n $, respectively. For the physical channels
only,
the RGM results looks quite different for the ${ t-p}$ $S$-matrix, even 
qualitatively. The standard calculation (curve b) displays already all the 
qualitative features of the $R$-matrix results. The model space of dimension 50
(curves c) reproduces in both cases the $R$-matrix results at low energies
perfectly, but crosses the imaginary axis below instead of above
the real axis (fig.\ \ref{nm2}) and vice versa (fig.\ \ref{nm3}). From these
figures it is obvious that small modifications in any of the approaches
might change the $S$-matrix a small amount, so that the behaviour near the
origin will also qualitatively agree. For the large model space, the
resonances are all shifted downward to 1.2, 5.8, and 10.9 MeV, thus
agreeing even better with the compilation \cite{We1} (see table \ref{levs}).

\vskip 0.2cm

\begin{figure} [t]
\centerline{\epsfxsize=10cm  \epsfbox{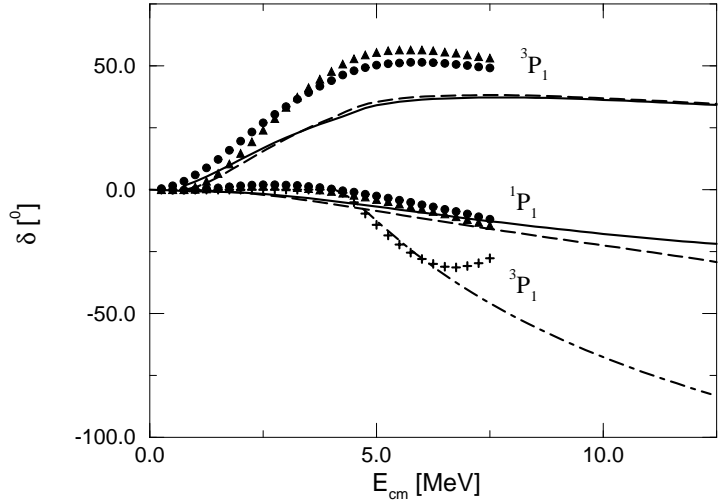}}

\caption{\label{em1} Same as fig.\ \protect \ref{np3},
but for the  $1^-$ phase shifts.}
\end{figure}

  The $1^-$ phase shifts are displayed in fig.\ \ref {em1}. The $R$-matrix
data yield
positive $^3P_1$ and negative $^1P_1$ [3+1] phases and also negative $d-d$
ones. The extracted level structure is quite rich \cite{We1}. The RGM 
calculation uses all physical channels and a few distortion channels. The
results compare favourably with the $R$-matrix ones, even though the $^3P_1$
phases are not quite as large. Allowing for a large number of distortion
channels, corresponding to those of the $0^-$ partial wave, the positive phases
grow up to 45 degrees, and so come into close agreement. From the eigenphase
shifts we deduce also four resonances (see table \ref{levs}), which
agree nicely in structure and position with the $R$-matrix ones. It should be
noted that the $^3P_1$ [3+1] resonances are strongly isospin mixed.

\vskip 0.2 cm

\begin{figure} [t]
\centerline{\epsfxsize=10cm  \epsfbox{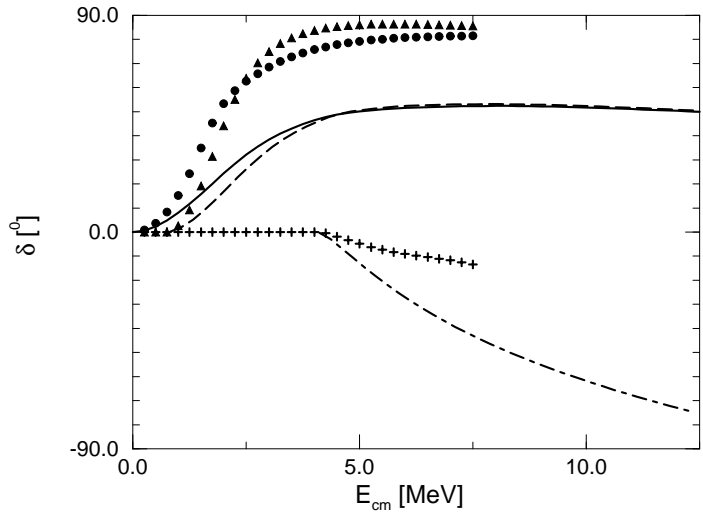}}

\caption{\label{zm1} Same as fig.\ \protect \ref{np3},
but for the  $2^-$ phase shifts.}
\end{figure}

  In the $2^-$ partial wave the $R$-matrix allows for $P$- and $F$-waves in all
configurations. In fig.\ \ref{zm1} the $P$-wave phase shifts are displayed.
The [3+1] phases reach about 90 degrees at rather low energies, which indicate
the existence of two lying resonances (see ref. \cite{We1} and table
\ref{levs}).
The ${ d-d}$ phases are negative as in the other two negative-parity 
partial waves. The RGM
calculation yields somewhat different results: the positive phase shifts
reach only half the $R$-matrix value, whereas the negative ${ d-d}$ phases
are about a factor two more negative. The extracted resonance positions are
still quite reasonable (see table \ref{levs}).

\vskip 0.2 cm

  Since the $R$-matrix analysis allows also for $F$-waves, there are also
results for the $3^-$ and $4^-$ partial waves. The $^1F_3$ phase shifts
are negative and reach -4 degrees at the highest energy. The $^3F_J$
phase shifts are all positive or tiny, i.e. in magnitude below 0.2 degrees.
The [3+1] $^3F_3$ phase shifts reach up to 7.0 degrees and the ${ d-d}$ $^3F_4$
one up to 4.8 degrees, all others are tiny. Resonant structures do not
appear. No RGM results exist till now for these partial waves.

\vskip 0.2cm

  For the negative parities the $R$-matrix and 
RGM phase shifts could be interpreted as
due to an effective strong tensor force for the [3+1] fragmentations. 
Whereas the $R$-matrix results for the ${ d-d}$
$F$-waves point more into the direction of
a stronger spin-orbit force.

\vskip 0.2cm

  The high lying resonances of ${ d-d}$ structure, which occur in almost all
partial waves deserve some further discussion. They are not related to any
circling of the corresponding $S$-matrix element as in the $0^+$ and $0^-$
partial waves (see figs.\ \ref{np4}, \ref{nm2}, and 
\ref{nm3}). They are due to the
prescription given in ref. \cite{We1}, where the ${ d-d}$ channel has a
channel radius of 7.0 fm. This large channel radius leads to rather large 
background
phase shifts, so that around 11 MeV the differences between the
eigenphases of the $S$-matrix and the background phase pass through ninety
degrees, the criterion we chose for defining the energetic position of a
resonance. So it might be that the position of 
all of these high energy $d-d$ resonances
is strongly influenced by the large value of the channel radius. Unfortunately
within the RGM we are not aware of any better criterion.
This might be the reason, why we find in the RGM calculation also in the $0^+$
and $1^+$ partial waves resonances around that energy, which have no
counterpart in the $R$-matrix analysis (cf.\ table \ref{levs}).

\section{Comparison with data}

\begin{figure}[t]
\centerline{\epsfxsize=12cm  \epsfbox{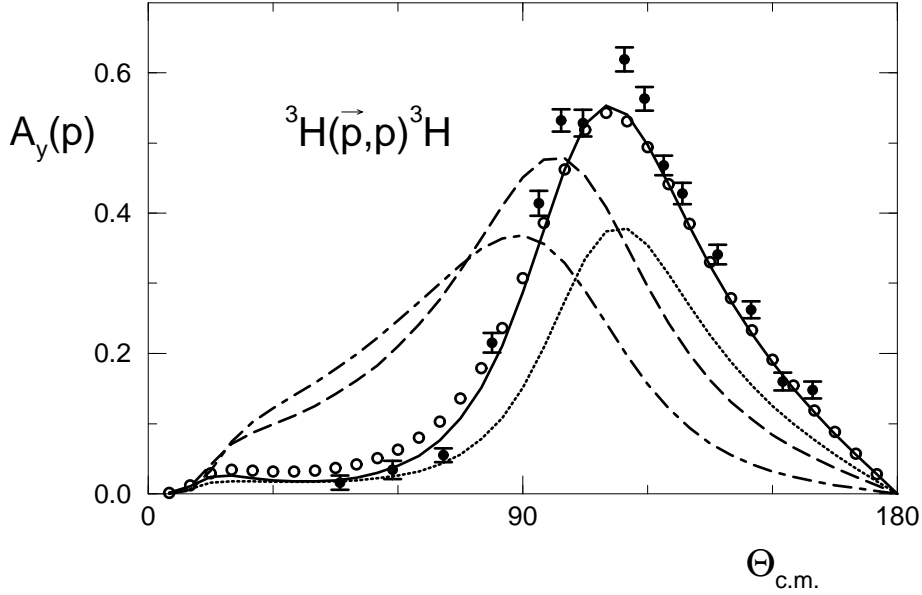}}
\caption{\label{tptpap} Proton analysing power of the reaction $^3{\rm H}
 (p,p){^3{\rm H}}$
calculated for 3.0 MeV $E_{\rm cm}$.
The data are for 4.15 MeV protons from ref. \protect \cite {Erlangen}.
The full line represents the
$R$-matrix analysis, the dashed one the full calculation, the dot-dashed one
the small calculation, and the dotted one the semi-realistic calculation.
The open circles denote
the full calculation with the $^3P_2$ matrix element replaced by the 
corresponding $R$-matrix one.}
\end{figure}

  In this chapter we compare the various calculations with a wide variety of
experimental data. We present figures for all possible elastic scatterings
and reactions. Out of the large amount of data, we have selected those energies
for which several observables are measured, if possible by different groups.
Most of the data presented here are in the data set used to determine the
$R$-matrix parameters. We will point out those that are not included in the
set. We present results for the $R$-matrix analysis together with RGM
calculations.
In order to demonstrate the effects of enlarging the model spaces in the
calculation we display curves for the largest model space used, denoted by
``full'', and for the negative parities for the physical model space only,
denoted by ``small''. To demonstrate how the calculation using a realistic
nucleon-nucleon interaction differs from previous calculations, we display
also the results from a previous calculation \cite {Ho1}, denoted by
``semi-realistic''. Since the $R$-matrix analysis uses most of the data as input
values, we consider the $S$-matrix elements determined in this analysis as
experimentally determined, despite the fact that the analyis has not yet fully
converged. We emphasize the differences  of the $S$-matrix elements
originating from the $R$-matrix analysis and the full RGM calculation and
present the results when we change the value of a specific matrix element from
its RGM value to its $R$-matrix value in order to demonstrate that it is just
this single matrix element that does not allow to reproduce the data. At the
end of this chapter we discuss the various differences and their possible
origin. It should be noted that the partial wave analysis  of the RGM
calculation and the $R$-matrix differ. The $R$-matrix analysis contains 
additionally the
positive-parity contributions for $J^{\pi} = 3^+$ and $4^+$ and all the
negative-parity $F$-wave contributions.
We will point out in the following in which reactions these differences
 play a substantial role. The
meaning of the lines in all figures is the same as given in fig.\ \ref{tptpap}.

\begin{figure}[t]
\centerline{\epsfxsize=12cm  \epsfbox{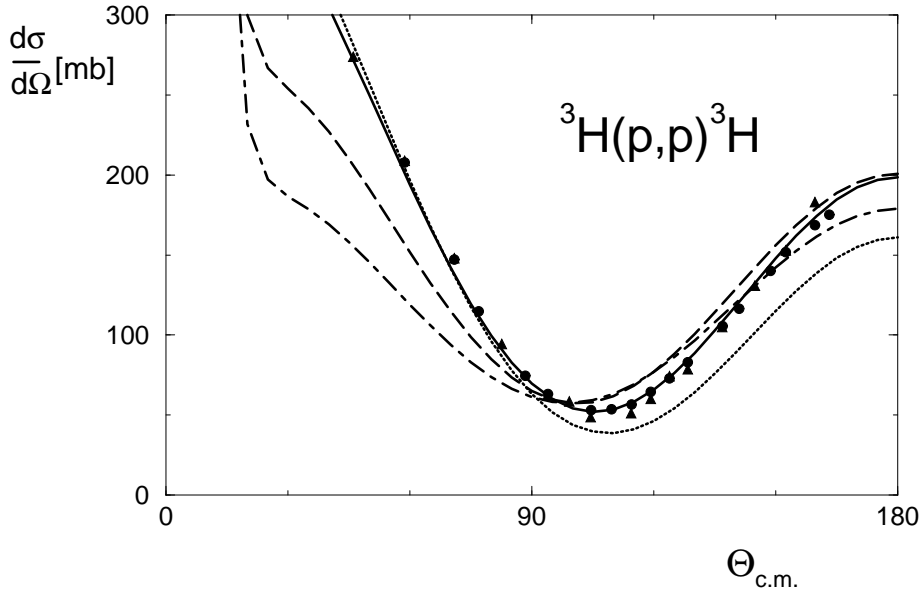}}
\caption{\label{tptpx} Differential elastic proton-triton
cross section calculated at 3.0 MeV $E_{\rm cm}$.
The data are for 4.15 MeV protons from ref. \protect \cite {Deto} (dots) and 
from ref. \protect \cite {Erlangen} (triangles).
The full line represents the
$R$-matrix analysis, the dashed one the full calculation, the dot-dashed one
the small calculation, and the dotted one the semi-realistic calculation.
The errors in the data are of the size of the symbols.}
\end{figure}
\vskip 0.2cm

\vskip 0.2cm

\vskip 0.2cm

  The various reactions are presented in the order of their thresholds,
starting with triton-proton elastic scattering. Here
differential cross section  and analysing power measurements for the proton
and also for the triton exist around 4 MeV proton energy. In fig.\ \ref{tptpx}
we compare the differential cross section data with the various calculations.

\begin{figure}[t]
\centerline{\epsfxsize=12cm  \epsfbox{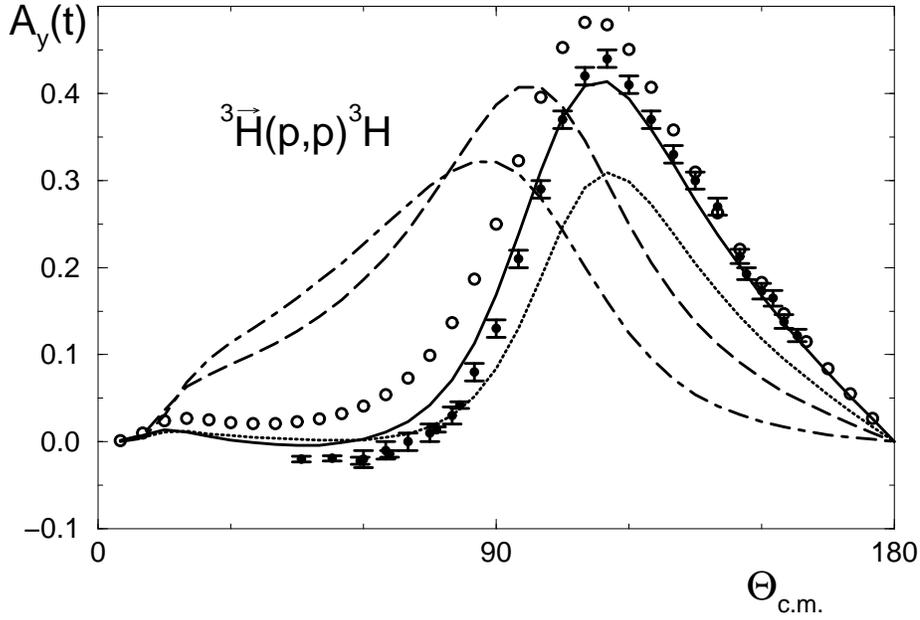}}
\caption{\label{tptpat} Triton analysing power of the reaction $^3{\rm H}
(p,p){^3{\rm H}}$
calculated 
for 3.0 MeV $E_{\rm cm}$.
The data are for 4.28 MeV protons from ref. \protect \cite {LA}. The labeling is
the same as in fig.\ \protect \ref{tptpap}.}
\end{figure}

The $R$-matrix analysis is performed in 250-keV steps of the center-of-mass
energy starting from the triton-proton threshold. Since there are no rapid
changes with energy of the observables presented, we use the results
of the calculations for the nearest energy to that of the data. The error
introduced by this procedure is usually well within the experimental error
bars. The RGM calculations are also in 250 keV steps of center-of-mass energy,
but the calculated thresholds are  slightly different (see table
\ref{thres}).
From fig.\ \ref{tptpx} we see that the $R$-matrix analysis reproduces the
data quite well. This demonstrates how good the fit to the data is in 
general. The
RGM calculation, on the other hand, reproduces the data reasonably well,
taking into account that the calculations contain no adjustable parameters.
There are, however, marked deviations from the data. The full calculation
using the Bonn potential describes the data much better than the small
calculation, but the deviations for forward angles are appreciable.

\begin{figure}[t]
\centerline{\epsfxsize=12cm  \epsfbox{tphnx.epsi}}
\caption{\label{tphnx} Differential cross section of the reaction 
$ {\rm ^3H}(p,n) ^3{\rm He}$ calculated for $E_{cm}=3.0$ MeV.
The data are for 4.101 MeV protons from ref. \protect \cite {Perry}.
The labeling is as in fig.\ \protect \ref{tptpap}.}
\end{figure}

The calculation using the semi-realistic potential describes the forward data
very well, but deviates strongly for larger angles. Since we know the
$S$-matrix elements from the calculations and the $R$-matrix analysis
themselves, we can
try to trace the differences to specific $S$-matrix elements. The full
calculation differs from the $R$-matrix analysis in the $^3P_2$ matrix element.
Note that we represent all $S$-matrix elements in the form
$S_{kl} = \eta_{kl} e^{2 i \delta_{kl}}$
with phase shifts $\delta_{kl}$ and channel couplings $\eta_{kl}$. In the
full calculation, $ \eta=0.86$ instead of 0.68 for the $R$-matrix analysis,
and the phase shift of 49 degrees
misses by 20 degrees the $R$-matrix result;
thus the coupling to the $^3{\rm He}-n$ channels is much
weaker. When we use the $R$-matrix $^3P_2$ matrix element instead of the
calculated one, the data are very well reproduced. The results lie on
top of the $R$-matrix results, therefore we do not display them in fig.\
\ref{tptpx}. The semi-realistic
calculation yields a much better $^3P_2$ matrix element. Here the deviations
are mainly due to too strong a coupling for the $^3P_1$ matrix element.
\vskip 0.2cm

  Proton analysing power measurements at the same energy as the cross section
and nearby triton analysing power data are displayed in figs.\ \ref{tptpap}
and \ref{tptpat}, respectively. The $R$-matrix analysis again does an 
excellent job, whereas the RGM calculation reproduces the data only
qualitatively
(see figs.\ \ref{tptpap} and \ref{tptpat}). This is again
due to the differences in the $^3P_2$ matrix elements (see open circles
in figs.\ \ref{tptpap} and \ref{tptpat}), which reproduce the data much better.
Note, however, that all calculations yield an analysing power for the proton
and the triton that is markedly different in the maximum. This difference
is due to the coupling matrix element $^1P_1 \rightarrow{^3P_1}$, which is
different from zero due to the spin-orbit component of the potentials used.
Contrary to the nucleon-nucleon system, such singlet-triplet transitions are
not forbidden by the Pauli principle.

\vskip 0.2cm

  The differential cross section for the ${\rm ^3H}(p,n) {\rm ^3He}$ reaction
at almost the same energy is displayed
in fig.\ \ref{tphnx}. Whereas the $R$-matrix reproduces the data quite well,
the RGM calculation yields only qualitative agreement with the data, with
the increased model space yielding again better agreement. The 
differences can again be traced to essentially a single matrix element, the 
already known 
$^3P_2$ one. For the Bonn potential it reaches only two thirds of the magnitude
of the $R$-matrix one, and its relative phase to the $^1S_0$ matrix element
is only 60 degrees instead of 80. This just demonstrates the unitarity
of the $S$-matrix, i.e. the missing coupling in the elastic triton-proton
channel now shows up in the 
missing strength going from the proton to the neutron channel. 
All other large matrix elements agree
within a few percent in modulus and phase.
Changing again in the full calculation the
$^3P_2$ matrix element to its $R$-matrix value reproduces the data almost
perfectly (cf. fig.\ \ref{tphnx}). Except in the minimum, the modified RGM
agrees with the $R$-matrix result.  For the semi-realistic potential,
the $^3P_2$ matrix element overshoots the $R$-matrix one by 25 percent and the
$^3P_1$ one by even 50 percent.

\vskip 0.2cm

\begin{figure}[t]
\centerline{\epsfxsize=12cm \epsfbox{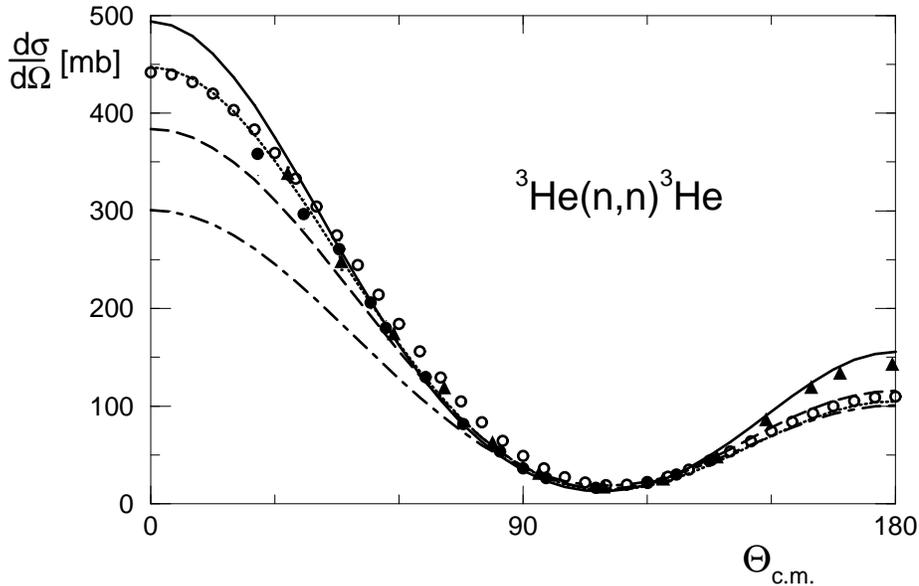}}
\caption{\label{hnhnx} Differential cross section calculated 
for 8MeV neutrons off $^3{\rm He}$.
The data are for 7.9 MeV neutrons from ref. \protect \cite {Drosg} (dots) and 
from ref. \protect \cite {Klages} (triangles) for 8 MeV.
The labeling is as in fig.\ \protect \ref{tptpap}.}
\end{figure}

\begin{figure}[t]
\centerline{\epsfxsize=12cm \epsfbox{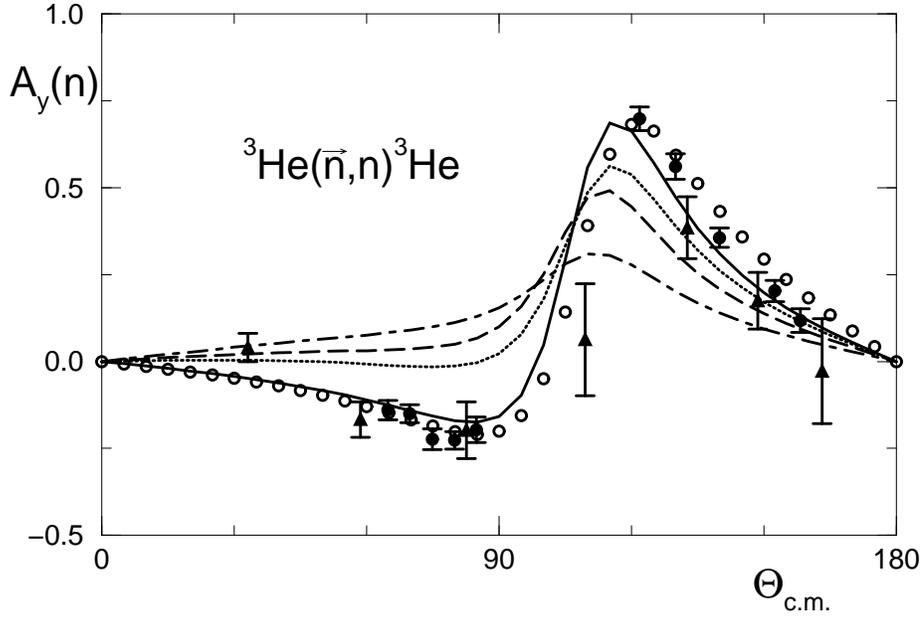}}
\caption{\label{hnhnap} Neutron analysing power for the scattering of 8 MeV
neutrons off $^3{\rm He}$. Data are 
from ref. \protect \cite{Drigo} (triangles) and from ref. \protect \cite {Lisow}
(dots). The labeling is as in fig.\ \protect {\ref{tptpap}}.}
\end{figure}

  Elastic neutron scattering cross section and neutron analysing powers around
8 MeV neutron energy are displayed in fig.\ \ref{hnhnx} and fig.\ \ref{hnhnap},
respectively.
The cross section data are well reproduced not only by the $R$-matrix
analysis but also by the full and semi-realistic calculations.
Changing in the full calculation
again the $^3P_2$ matrix element to its $R$-matrix value brings the results on 
top of the semi-realistic cross section results.

\begin{figure}[t]
\centerline{\epsfxsize=12cm  \epsfbox{ddtpx.epsi}}
\caption{\label{ddtpx} Differential cross section for the reaction $^2{\rm H}
(d,p){\rm ^3H}$
calculated for  $E_{cm}=2.11$ MeV.
 Data are for 4.0 MeV deuterons
from ref. \protect \cite{grueb} (triangles) and from ref. \protect \cite
{schulte} (dots). The labeling is as in fig.\ \protect {\ref{tptpap}}.}
\end{figure}

\begin{figure}[t]
\centerline{\epsfxsize=12cm  \epsfbox{ddtpt11.epsi}}
\caption{\label{ddtpt1} The analysing power $iT_{11}$
for the reaction $^2{\rm H}(d,p){\rm ^3H}$
calculated for  $E_{cm}=2.11$ MeV.
 Data are for 4.0 MeV deuterons
from ref. \protect \cite{grueb}.The labeling is the same as in fig.\ 
\protect \ref{tptpap}.}
\end{figure}

\vskip 0.2cm

The neutron
analysing power, however, is well reproduced by the $R$-matrix, but only 
qualitatively by the RGM-calculations.
The height at the maximum analysing power
around 115 degrees is controlled by the phase of the $^3P_2$ matrix element,
since the modulus of the matrix elements turns out to be quite similar for 
all calculations.
In the full calculation the phase shift is about 25 degrees below the
$R$-matrix analysis (see fig.\ \ref{zm1}). Increasing the phase of the RGM
matrix element by this amount, as was done above for the cross section,
reproduces also the analysing power data perfectly.
Also the minimum around 90 degrees is well reproduced for this choice.
It should be mentioned, however, that the coupling matrix element
$^1P_1 \rightarrow { ^3P_1}$
is only about half the $R$-matrix one, thus leading to a much
smaller difference between the neutron and $^3{\rm He}$ polarisations (compare
with the discussion of the charge conjugate scattering ${\rm ^3H}(p,p){\rm ^3H}$ above).
There are, however, no $^3{\rm He}$ polarisation data available, hence one
cannot
decide about which calculation is correct. All the above elastic neutron
scattering data are not included in the $R$-matrix fit, hence, we consider
the calculations to be predictions.

\vskip 0.2cm

  Let us now come to the deuteron induced reactions. In fig.\ \ref {ddtpx}
the ${\rm ^2H}(d,p){\rm ^3H}$ differential cross section is displayed.
The data are well
reproduced by all calculations. Here for the first time the small calculation
reproduces the data even better than the full calculation. The semi-realistic 
calculation overshoots the data a bit. At the forward (and
backward) angles, the $F$-waves, which are missing in the RGM calculations, 
contribute about 50 percent to the cross section
in the $R$-matrix analysis and indicate that even G-waves may be necessary.

\vskip 0.2cm

\begin{figure}[t]
\centerline{\epsfxsize=12cm  \epsfbox{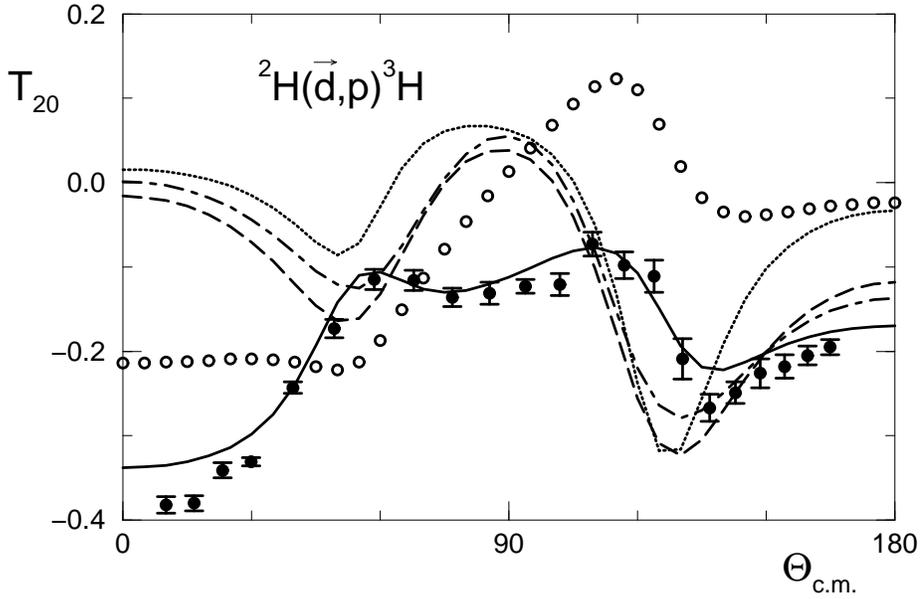}}
\caption{\label{ddtpt20}  Same as fig.\ \protect \ref{ddtpt1},
but for the analysing
power $T_{20}$.}
\end{figure}

\begin{figure}[t]
\centerline{\epsfxsize=12cm  \epsfbox{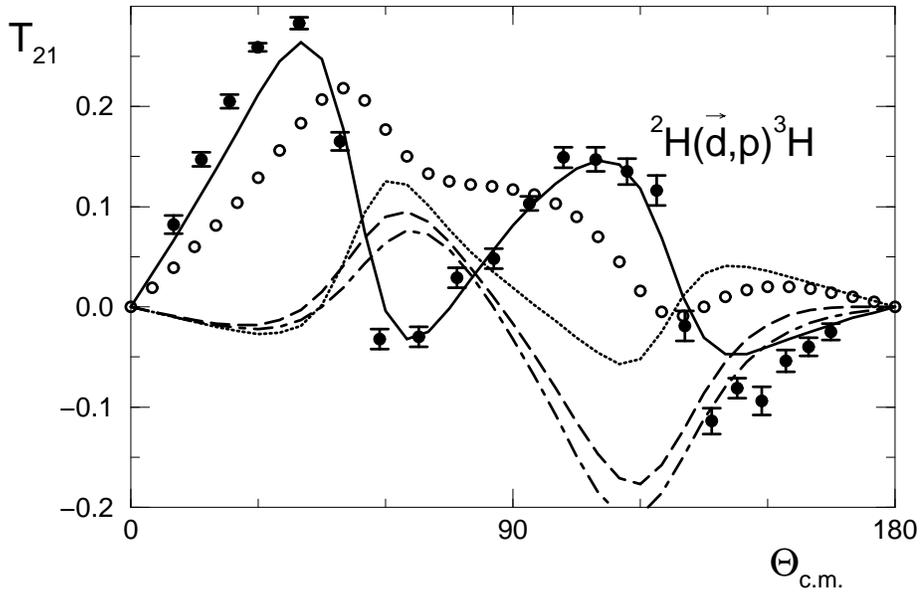}}
\caption{\label{ddtpt21}  Same as fig.\ \protect 
\ref{ddtpt1}, but for the analysing
power $T_{21}$.}
\end{figure}

\begin{figure}[t]
\centerline{\epsfxsize=12cm \epsfbox{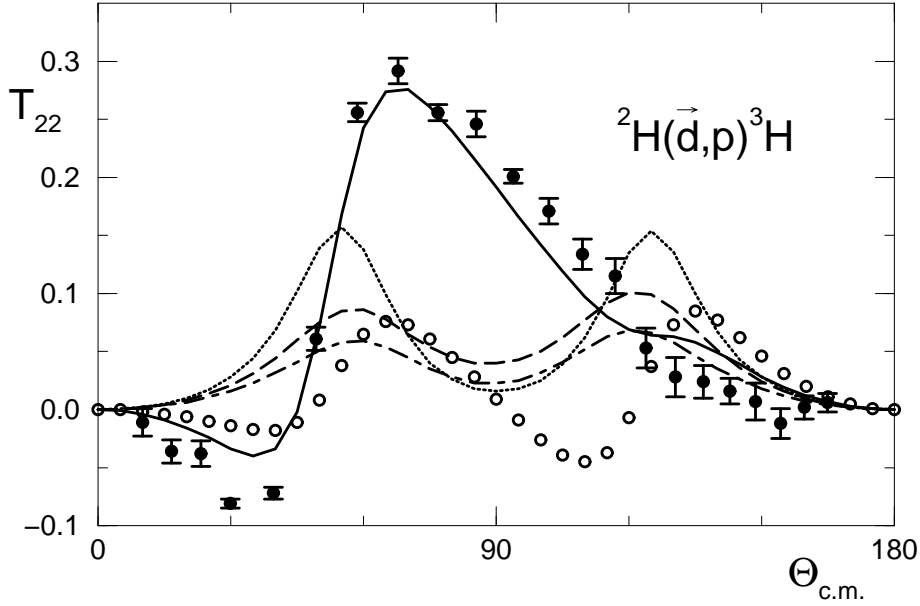}}
\caption{\label{ddtpt22}  Same as fig.\ \protect 
\ref{ddtpt1}, but for the analysing
power $T_{22}$.}
\end{figure}

In figs.\ \ref {ddtpt1}, \ref{ddtpt20}, \ref{ddtpt21}, and \ref{ddtpt22}
the deuteron analysing powers are displayed. Whereas the $R$-matrix analysis
describes the data nicely, the RGM calculations fail completely.
In the $R$-matrix analysis the dominant matrix elements are the $^3P_1$, $^1D_2$,
and the $^5S_2 \rightarrow{^3D_2}$
matrix elements, but also quite a number of others must not be neglected,
like the $^1S_0$ and the $^5S_2 \rightarrow{^1D_2}$.
The coupling matrix element $^5S_2 \rightarrow{^3D_2}$
determines essentially the form of the angular distribution of all polarisation
observables. The $R$-matrix analysis finds this matrix element to half of the
leading $^3P_1$ one and the relative phase shift to be +60 degrees. The 
full RGM
calculation yields, however, only a quarter of the leading matrix element
and the relative phase shift of -33 degrees, i.e. the opposite sign.
The $^1S_0$ and $^3P_2$ overshoot the $R$-matrix result by a factor 2 to 3. 
The other
RGM calculations give similar results. Changing the RGM coupling matrix element
$^5S_2 \rightarrow{^3D_2}$
to the $R$-matrix value yields reasonable agreement for the vector polarisation 
data (cf.\ fig.\ \ref{ddtpt1}) and improves the angular form of the tensor
analysing powers (cf.\ figs.\ \ref{ddtpt20}, \ref{ddtpt21}, and \ref{ddtpt22}).
For the forward angles the calculation comes now much closer to the data, but 
the shape of the angular distribution does not match. The reason for this
discrepancy can be found in the interference terms
of the large and the smaller matrix elements.
The $R$-matrix analysis indicates that
for a detailed description of these data, the $F$-wave contributions
are necessary.
The modified cross section turns out to be slightly too large.

\vskip 0.2cm

  For the ${\rm ^2H}(d,n){\rm ^3He}$ reaction, data exist for the same energy
as for the charge conjugate ${\rm ^2H}(d,p){\rm ^3H}$ reaction.
In fig.\ \ref {ddhnx} we compare the 
differential cross section data with the various calculations. The $R$-matrix
analysis again does a perfect job. Also the full RGM calculation reproduces
the data nicely, whereas the semi-realistic one overshoots the data
appreciably.
The dominant matrix elements are again $^3P_1$, $^1D_2$, and 
$^5S_2 \rightarrow {^3D_2}$ in the $R$-matrix analysis, but also the $^1S_0$,
$^3P_2$, and $^5S_2 \rightarrow {^1D_2}$ being of equal magnitude, must not be
neglected. The full RGM calculation yields the two dominant matrix elements
similarly, but the coupling matrix elements $^5S_2 \rightarrow {^3D_2}$
and also $^5S_2 \rightarrow {^1D_2}$ are only half of the $R$-matrix value.
On the other hand, the $^1S_0$ and $^3P_2$ ones are doubled, as is the case
of the charge conjugate reaction.

\begin{figure}[t]
\centerline{\epsfxsize=12cm \epsfbox{ddhnx.epsi}}
\caption{\label{ddhnx} Differential cross section for the reaction $^2{\rm H}
(d,n){\rm ^3He}$
calculated for  $E_{cm}=2.11$ MeV.
 Data are for 4.0 MeV deuterons from ref. \protect \cite {schulte}.
The labeling is as in fig.\ \protect {\ref{tptpap}}.}

\end{figure}

\begin{figure}[t]
\centerline{\epsfxsize=12cm \epsfbox{ddhnay.epsi}}
\caption{\label{ddhnay} The deuteron vector analysing power
for the reaction $^2{\rm H}(d,n){\rm ^3He}$ calculated for  $E_{cm}=2.11$ MeV.
 Data are for 4.0 MeV deuterons
from ref. \protect \cite{dries}.The labeling is the same as in fig.\ 
\protect \ref{tptpx}.}
\end{figure}

\begin{figure}[t]
\centerline{\epsfxsize=12cm \epsfbox{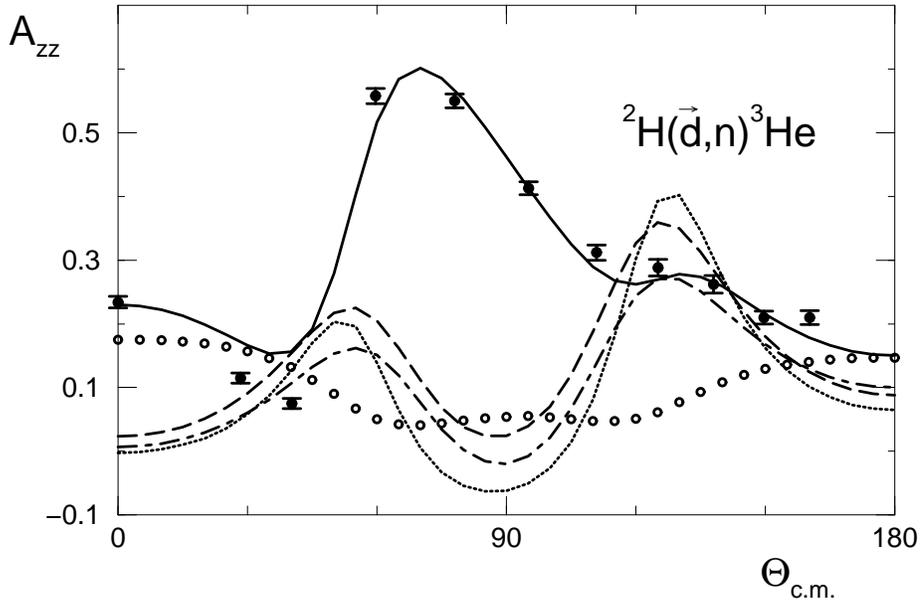}}
\caption{\label{ddhnazz}  Same as fig.\ \protect 
\ref{ddhnay}, but for the analysing
power $A_{zz}$.}
\end{figure}

  The Ohio-State 
group \cite{dries} has measured all deuteron polarisation observables
also for 4.0 MeV deuterons. 
In figs.\ \ref{ddhnay} and \ref{ddhnazz}, we present only the vector analysing
power and $A_{zz}$ data, because the overall behaviour is quite similar to the
charge conjugate reaction.

As in the charge conjugate reaction ${\rm ^2H}(d,p){\rm ^3H}$, only the 
$R$-matrix analysis reproduces the data, whereas all RGM calculations fail
completely. To find such a good agreement with data, the $^3F_3$ and $^3F_4$
matrix elements in the $R$-matrix are vital. Again the
$^5S_2 \rightarrow {^3D_2}$
matrix element is dominating the angular form of all analysing powers. Using
in the full RGM calculation the corresponding
$^5S_2 \rightarrow {^3D_2}$ matrix element from the $R$-matrix does not
reproduce the data much better. Here it
is also neccessary to modify the corresponding
$^5S_2 \rightarrow {^1D_2}$ matrix element. Then the cross section is not 
changed at all, the vector analysing power is reproduced reasonably well,
but for the description of the tensor polarisations, not too much is gained
(cf.\ figs.\ \ref{ddhnay} and \ref{ddhnazz}). Note that in the RGM no
$F$-waves are included.

\vskip 0.2cm

\begin{figure}[t]
\centerline{\epsfxsize=12cm  \epsfbox{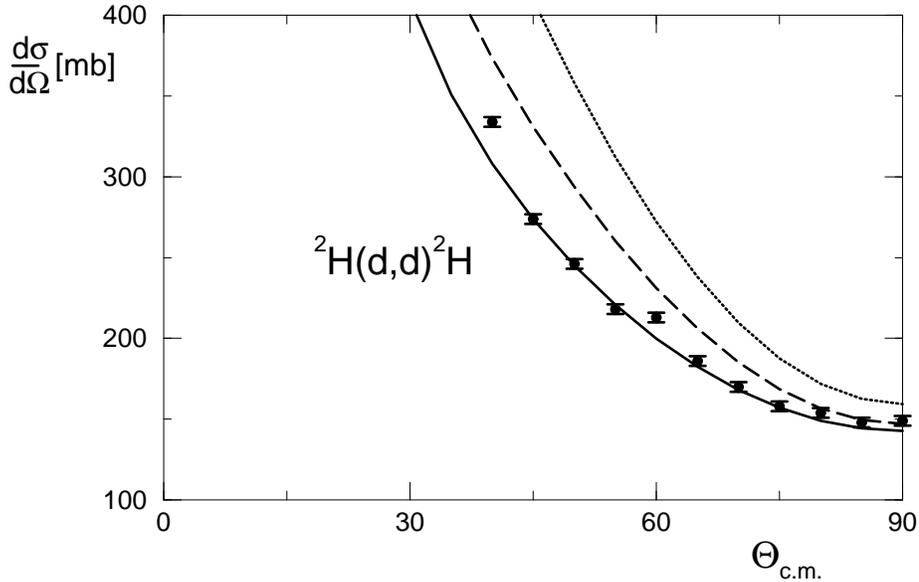}}
\caption{\label{ddddx} Differential cross section for elastic deuteron-deuteron
scattering calculated for  $E_{cm}=3.11$ MeV.
 Data are for 6.0 MeV deuterons from ref. \protect \cite{ddc}.
The labeling is as in fig.\ \protect {\ref{tptpx}}.}
\end{figure}

  Let us now discuss the last two-fragment process, elastic deuteron-deuteron
scattering. Because of the identical bosons in the entrance and exit channel,
the vector polarisation and $T_{21}$ are antisymmetric and all other
observables are symmetric about
90 degrees. Since the data are usually converted into the forward hemisphere
we display this part only in the following figures.
In fig.\ \ref{ddddx} we present rather old data \cite {ddc} for the
differential cross section together with the calculations. The $R$-matrix
analysis reproduces the data very well. The two realistic RGM calculations
lying on top of each other also reproduce the data reasonably 
well, the semi-realistic
calculation, however, overshoots the data appreciably.

\vskip 0.2cm
\begin{figure}[t]
\centerline{\epsfxsize=12cm  \epsfbox{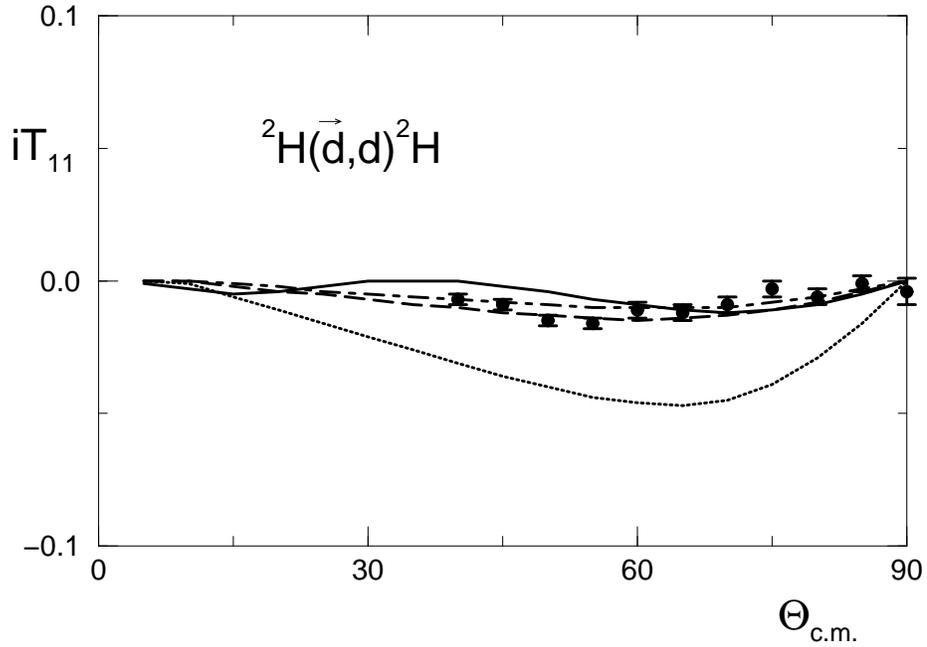}}
\caption{\label{ddddt1} Vector analysing power $iT_{11}$ 
for elastic deuteron-deuteron scattering calculated for  $E_{cm}=3.11$ MeV.
 Data are for 6.0 MeV deuterons from ref. \protect \cite{grueb73}.
The labeling is as in fig.\ \protect {\ref{tptpx}}.}
\end{figure}

\begin{figure}[t]
\centerline{\epsfxsize=12cm  \epsfbox{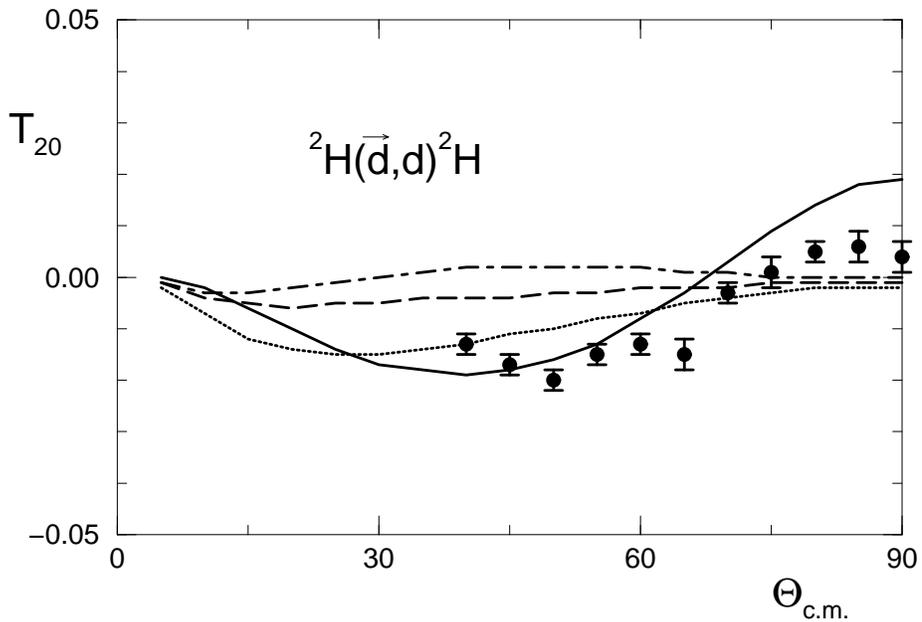}}
\caption{\label{ddddt20} Same as fig.\ \protect {\ref{ddddt1}} but for the
tensor analysing power $T_{20}$.}
\end{figure}

\begin{figure}[t]
\centerline{\epsfxsize=12cm  \epsfbox{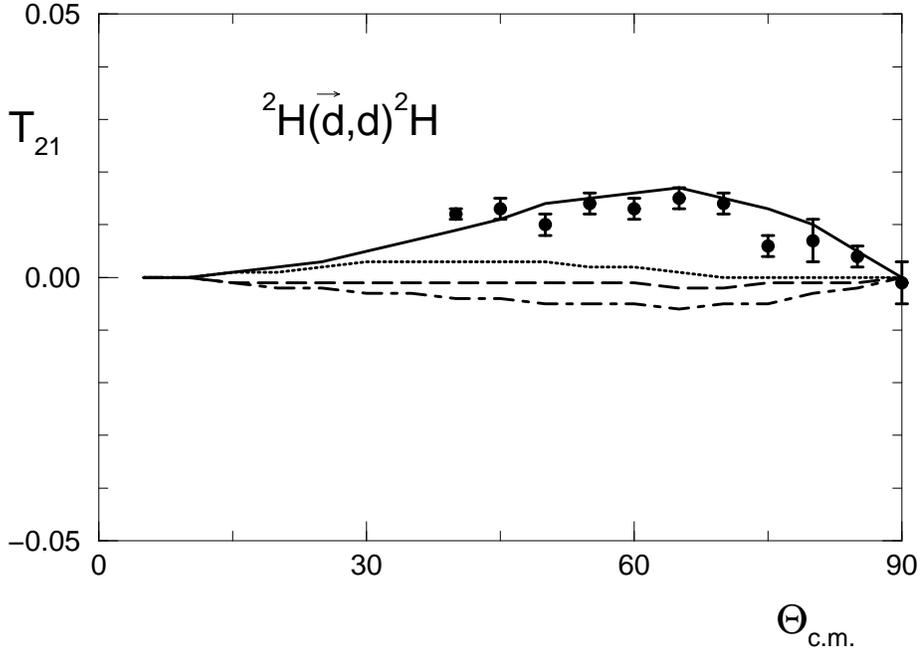}}
\caption{\label{ddddt21} Same as fig.\ \protect {\ref{ddddt1}}, but for the
tensor analysing power $T_{21}$.}
\end{figure}

\begin{figure}[t]
\centerline{\epsfxsize=12cm  \epsfbox{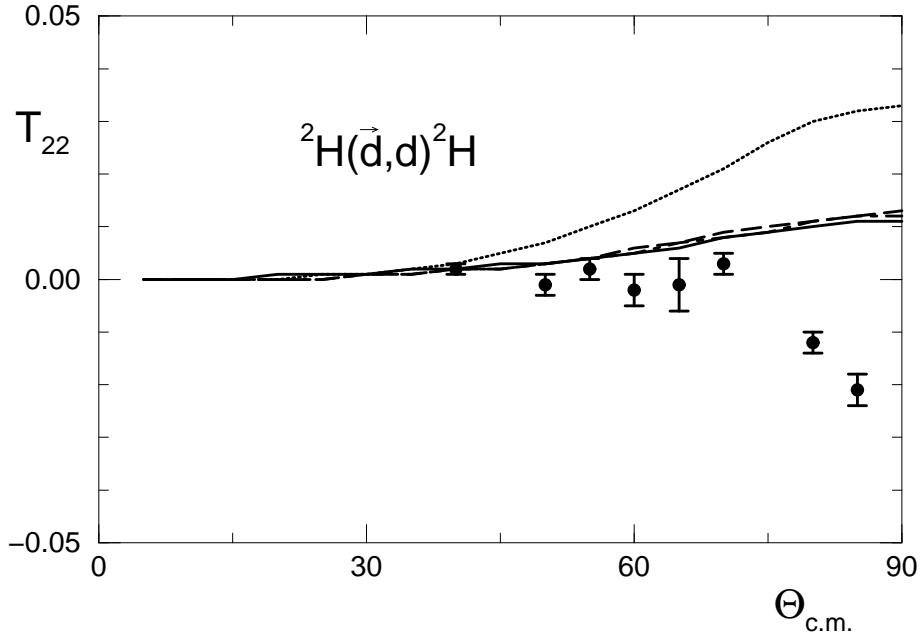}}
\caption{\label{ddddt22} Same as fig.\ \protect {\ref{ddddt1}}, but for the
tensor analysing power $T_{22}$.}
\end{figure}

  The analysing powers have been measured by the Z\"urich group \cite {grueb73}.
In figs.\ \ref{ddddt1},
\ref{ddddt20}, \ref{ddddt21}, and \ref{ddddt22} we compare the data with
various calculations. All polarisations are quite small,
of the order of one or two percent. Note the large scale on the figures.
Except for the tensor polarisation $T_{22}$, where all calculations fail, the
$R$-matrix analysis reproduces the data well. Considering the small values,
also the full calculation reproduces the data reasonably well. Since the
polarisations
are so small, it is useless to search for the matrix element(s), which
dominate the structure of the polarisations. It has to be mentioned,
however, that the inclusion of these tiny polarisations into the $R$-matrix
analysis changed the sign in the $^3D_2$ triton-proton and neutron-$^3{\rm He}$
elastic matrix elements (see the above discussion following fig.\ \ref{zp1}).

\vskip 0.2cm

  Considering all the figures displaying a comparison of data and calculations
together, it is obvious that the parameter-free calculations reproduce the
differential cross sections rather well, whereas the polarisations are
sometimes grossly missed.
The differences in the description of the data between the $R$-matrix 
analysis and the RGM-calculations are concentrated in a few $S$-matrix elements.
In order to reproduce the data the calculated $^3P_2$ matrix elements for all
channels and the $^5S_2 \rightarrow {^3D_2}$ and the $^5S_2 \rightarrow {^1D_2}$
matrix element in the deuteron-deuteron fusion reaction have to be changed.
In the $^3P_2$ one, the coupling to the $F$-waves cannot be totally neglected.
So in both cases the tensor force is crucial for the determination of these
matrix elements, because the coupling is to channels with a different
orbital angular momentum. As discussed in the previous chapter, the $R$-matrix
analysis indicates a rather strong tensor force. 
It might therefore be that the Bonn potential is 
the origin for the discrepancies of the RGM calculation with data. That
the tensor component in the Bonn potential is rather weak has been pointed out
many times, e.g. \cite {bonntensor}.

\section{Conclusion and Outlook}

  We performed microscopic multi-channel calculations employing an $r$-space
version of the realistic Bonn potential in the framework of the resonating
group model. This is a parameter-free calculation in a rather large model
space, chosen by physical arguments and restricted by computational
limitations. 
For the ground state it is essentially a variational calculation, with the
functional form of the wave function dictated by the following scattering
calculation. 

\vskip 0.2cm

  For the ground state energy we got almost the same results
as were obtained using variational wave functions \cite{Ca1} for the same
potential. The extracted point Coulomb form factor agrees nicely with that of
a variational calculation \cite{Schia} 
using another realistic nucleon-nucleon potential. 
Restricting the model space to those parts appropriate for bound state wave
functions, we could determine properties of the first excited state and also
the point Coulomb form factor. From the properties  we concluded that this
resonance is not a breathing mode and gave arguments why the energy of this
resonance in shell model calculations differs from that determined from
scattering experiments. Also the form factor deviates strongly from that of
the ground state.

\vskip 0.2cm

  From the scattering calculations we extracted phase shifts and demonstrated
that the parameter-free calculation yields not only qualitatively, but in
most cases even quantitatively, the same
answers as the $R$-matrix analysis using only data.
In some cases we showed that the apparent differences of the two methods are
caused by only minor variations of the $S$-matrix elements near the origin.
The other deviations appear in channels that can be strongly influenced by
the tensor-force terms in the nucleon-nucleon interaction. Since it is well
known that the potential we used has a rather weak tensor component
\cite{bonntensor}, we blame the potential for those differences.
Extracting resonance positions in an analogous way to the recent compilation
\cite{We1} yielded close agreement in the sequence, quantum numbers,
and structure of the resonances. Near the upper limit of the energy interval
considered, the difference between eigenphases of mainly deuteron-deuteron 
structure
and background phase shifts passed through ninety degrees, the accepted 
criterion for the position of
a resonance. In these cases, however, the interpretation as a resonance seems
to be doubtful.

\vskip 0.2cm

  Finally we compared the various calculations with data. All differential
cross sections were well reproduced, especially if we used the largest
model space. The polarisation observables, however, revealed the shortcomings
mentioned above. Only after modifying in most cases a single $S$-matrix
element, the agreement between calculation and data became satisfactory.
For the $ d-d $
fusion reaction, however, $F$-wave contributions that could not be taken
into account and the interference effects with smaller matrix elements
prevented a reasonable agreement.

\vskip 0.2cm

  Taking all this information together, the parameter-free microscopic
calculations using a realistic nucleon-nucleon force are very promising in
explaining and even predicting data. The shortcomings discussed above
indicate the direction of further studies. Increasing the model space brings
calculations and data in closer agreement (cf.\ figs.\ \ref{tptpx} to
\ref{ddddt22}), but the essential deviations between $R$-matrix analysis and
RGM calculations are untouched. Since we suspect the nucleon-nucleon potential
is responsible 
for these deviations, it is necessary to use another potential with a stronger
tensor component. Since the RGM requires the potential to be given in
$r$-space, we consider the Argonne AV14 \cite{av14} the proper choice,
especially because there exist many  calculations of ground state properties
for  light nuclei with  this potential. Furthermore, results for this two-body
potential supplemented by various three-body potentials are also 
available. Work
in this direction is under way.

\vskip 0.2cm

Acknowledgements: Both authors wish to acknowledge the hospitality of
various institutions, where part of the work was done: Los Alamos
National Laboratory, where all this work started quite some time ago,
while one of us (H. M. H.) 
was on sabbatical, the ECT*/Trento, where most of the manuscript
was written, and Triangle Universities Nuclear Laboratory, where final
preparation of the manuscript took place.
This work would not have been possible without the support in part by funds
from the U. S. Department of Energy Division of Nuclear Physics (G. M. H.)
and from the Deutsche Forschungsgemeinschaft, the BMFT and the BMBF (H. M. H.).

\begin{appendix}

\section
{Appendix 1: Wave functions of the fragments}

  The general structure of the wave functions is given in eqs. (1) and (2) 
(for further details see \cite {Lis,Ho2}). For the deuteron we have used 
the most simple wave function. Since the spin part is trivial, we give just the
orbital part.
\begin{equation}
\psi_D = \sum_{n=1}^3 c_n  e^{- \beta _n  /2 \; ( {\bf r}_1 -{\bf r}_2 )^2} \;
{\cal Y}_0 ( {\bf r}_1 -{\bf r}_2 )  \;
+\sum_{n=4}^5 c_n  e^{- \beta _n  /2 \; ( {\bf r}_1 -{\bf r}_2 )^2} \;
{\cal Y}_2 ( {\bf r}_1 -{\bf r}_2 ) 
\end{equation}

Here $\cal Y$ denote solid spherical harmonics \cite {Ed} .
The width parameters $\beta _n$ and the coefficients $c_n$ for the deuteron and
the singlet deuteron $ \dbar $ are given in table \ref{dwi}.

\begin{table}[b] \caption{\label{dwi} Deuteron and singlet deuteron parameters}
\vskip0.2cm
\centering
 \begin{tabular}{c|c|c|c|c|c}
n & 1 & 2 & 3 & 4 & 5 \\ \hline
$\beta _n$ & 5.8359 & 0.5201 & 0.0684 & 1.625 & 0.4339 \\
$c_d$ & 6.239420 & -6.659597 & -2.370518 & -6.685710 & -0.7940629 \\
$c_{\bar d}$ & -4.491821 & 3.059319 & 3.018020 & 0 & 0 \\
\end{tabular}
\end{table}

  The spatial part of the three-nucleon wave functions is of the form
\begin{equation}
\psi_{3N} = \sum_{l_1,l_2,L,n,m} \; c_{n,m}^{\left[ l_1,l_2 \right ] L} \;
e^{ - \beta_n s_1^2 } \; e^{ - \beta_m s_2^2 } \;
{\cal{Y}}_{l_1}({\bf s}_1) \; {\cal{Y}}_{l_2}({\bf s}_2) \; \phi^S
\end{equation}

where $\phi^S$ denotes the appropriate spin functions of each term in the sum.
The Jacobian coordinates are given by
\begin{equation}
{\bf s}_1 = \frac {1}{\sqrt{2}}({\bf r}_1 - {\bf r}_2 ) \; \; \mbox{ and }
\; \;  
{\bf s}_2 = \frac {1}{\sqrt{6}}({\bf r}_1 + {\bf r}_2 ) \;- \sqrt{\frac{2}{3}}
\; {\bf r}_3
\end{equation}
 
The width parameters $\beta_n$ and  $\beta_m$ for triton and $^3{\rm He}$ 
are given in table \ref{3hwi}.
\begin{table}[ht] \caption{\label{3hwi} Triton  and $^3{\rm He}$ width
parameters}
\vskip0.2cm
\centering
 \begin{tabular}{c|c|c|c|c|c|c}
k & 1 & 2 & 3  &  4 &5 &6 \\ \hline
$\beta _n^{^3{\rm H}}$ &   4.972 & 0.8659  & 0.1531 & 8.665 &1.8056 &0.5414 \\
$\beta _m^{^3{\rm H}}$ &  3.293 &0.3972  &  0.07513 & 1.626 &0.4745 & 0.09417\\
$\beta _n^{^3{\rm He}}$ & 4.8887 &0.8286  &0.1493 & 8.354 & 1.939 & 0.61723 \\
$\beta _m^{^3{\rm He}}$ & 3.616 & 0.3822 & 0.06763 &1.546 & 0.4668 & 0.09258 \\
\end{tabular}
\end{table}
The first three parameters belong to the structures containing no $D$-waves
and the last three width parameters belong to those containing at least one
$D$-wave.

\section{Appendix 2: Minimal $^4{\rm \bf He}$ Wave Function}

  Since the full bound state wave function of the $^4{\rm He}$ ground state
contains 227 different configurations, it is too complicated to be given in
detail. To give an impression about the dominant structures of the
ground state wave function, we will describe in some detail the minimal
wave function of table \ref{qb}. It consists of 20 configurations, where 
11 originate from the physical channels, which have been resolved into 
its constituents, and 9 are former distortion channels. All of them have
orbital angular momentum zero on the relative coordinate between the fragments,
but as can be seen in table \ref{qb}, all spin values are present.
For the relative motion we choose 5 width parameters, used also for all
distortion channels irrespective of the fragmentation, with the values:
2.433, 0.8214, 0.2987, 0.1185, 0.05001 in units of ${\rm fm}^{-2}$. For the
internal width parameters of the fragments the parameters given in tables
\ref{dwi} and \ref{3hwi} are used.

\vskip 0.2cm
  The dominant $S=0$ contribution contains 11 configurations: In the
triton-proton fragmentation, the three containing no internal angular momentum
from the physical channel and also the combination with the proton-neutron
pair inside the triton coupled to spin=1, but with the coefficients for the
various internal width parameters taken for the third exited state. The
$^3{\rm He} - n$
fragmentation contains the analogous three contributions from the
physical channel. From the distortion channels we also find the vector
with the internal proton-neutron pair of $S = 1$, but here the second and third
excited states appear with large expansion coefficients, indicating that
together
with some other state they are numerically almost linearly dependent. For the
deuteron-deuteron fragmentation none of the physical components contribute.
The $S$-wave part of the deuteron, together with the $S$-wave part of 
the first and second excited states of the deuteron, form the remaining two
channels. The overlap of the components originating from the physical channels
with the total wave function is in excess of 90 percent. The
overlap of the configurations containing excited fragment wave functions
is only about 84-85 percent, whereas the one containing the first excited state
of the deuteron has less than 10 percent.

\vskip 0.2cm
  The next important $S=2$ contribution contains 8 configurations: For the
triton-proton fragmentation only the two configurations with one
internal angular momentum being two, the other zero, originating from
the physical channel, contribute. For the $^3{\rm He} - n$
fragmentation, the
configuration with $P$-waves on both internal coordinates and that with a
$D$-wave between the internal neutron-proton pair contribute from the physical
channel. From the distortion channels the two configurations with one internal
$D$-wave, with the coefficients from the first excited state contribute.
For the deuteron-deuteron fragmentation, the $S$-wave part of the deuteron
and the $D$-wave of the deuteron and the first excited state of the deuteron
form the two contributing channels. The overlap of the various components
with the total wave function varies between 3 percent for the $d-d$ distortion
channel and 27 percent for the $^3{\rm He}$  $D$-wave physical channel.

\vskip 0.2cm
  The $S=1$ component is formed from the $D$-wave part of the deuteron and
the $D$-wave of the first excited state, coupled to spin equal one.

\end{appendix}


\begin{thebibliography}{99}
\bibitem {We1} D.R. Tilley, H.R. Weller and G.M. Hale, Nucl. Phys.  {\bf
A541} (1992) 1
\bibitem {Ha1} G.M. Hale, D.C. Dodder and K. Witte (unpublished)

\bibitem {Ca1} J. Carlson, Phys. Rev. {\bf C38} (1988) 1879
\bibitem {Ho1} H.M. Hofmann, W. Zahn and H. St\"owe, Nucl. Phys. {\bf A357}
(1981) 139
\bibitem {Reto} M. Baumgartner, H.P. Gubler, M. Heller, G R. Plattner, W.
Roser and I. Sick, Nucl. Phys. {\bf A368} (1981) 189
\bibitem {Fi1} S. Fiarman and W.E. Meyerhof, Nucl. Phys. {\bf A206} (1973) 1
\bibitem {Mac} R. Machleidt, K. Holinde and Ch. Elster, Phys. Rep. {\bf C149}
(1987) 1
\bibitem {Lis} H.M. Hofmann, in: Models and Methods in Few-Body Physics,
eds. L. S. Ferreira et al., Lecture Notes in Physics {\bf 273}, Springer,
Heidelberg 1987
\bibitem {Wil} K. Wildermuth and Y.C. Tang, A Unified Theory of the Nucleus,
Vieweg, Braunschweig 1977
\bibitem {Tang} Y. C. Tang, in: Topics in Nuclear Physics, Lecture
 Notes in Physics {\bf 145}, Springer, Heidelberg 1981
\bibitem {Kel} H. Kellermann, H.M. Hofmann and Ch. Elster, Few-Body Sys.
{\bf 7} (1989) 31
\bibitem {Eik} H. Eikemeier and H.H. Hackenbroich, Nucl. Phys. {\bf
 A169} (1971) 407
\bibitem {Ho2} H.M. Hofmann, in Finite Systems and Multiparticle Dynamics,
vol. 3, D. A. Micha and F. S. Levin eds., Plenum to be published
\bibitem {Kohn} W. Kohn, Phys. Rev. {\bf 74} (1948) 1763
\bibitem {Ha2} G.M. Hale et al. (1989) (unpublished)
\bibitem {Ha3} G.M. Hale et al. (1983) (unpublished)
\bibitem {HH} W. Heitler, The Quantum Theory of Radiation, Oxford
University Press 1954
\bibitem {Wig} E.P. Wigner and L. Eisenbud, Phys. Rev. {\bf 72} (1947) 29;
A.M. Lane and R.G. Thomas, Rev. Mod. Phys. {\bf 30} (1958) 257
\bibitem {Ha4} G.M. Hale, R.E. Brown and N. Jarmie, Phys. Rev. Lett.
 {\bf 59} (1987) 763
\bibitem {Ko} P.E. Koehler et al., Phys. Rev. {\bf C37} (1988) 917

\bibitem {Tay} R.J. Eden and J.R. Taylor, Phys. Rev. {\bf B133} (1964)
 1575
\bibitem {Ed} A.R. Edmonds, Angular Momentum in Quantum Mechanics,
Princeton University Press 1960
\bibitem {Ceu} R. Ceulener, P. Vandepeutte and C. Semay, Phys. Rev.
 {\bf C38} (1988) 2335
\bibitem {Schia} R. Schiavilla, V.R. Pandharipande and D.O. Riska,
 Phys. Rev. {\bf C41} (1990) 309
\bibitem {Ar} R.G. Arnold et al., Phys. Rev. Lett. {\bf 40} (1978) 1429
\bibitem {Fro} R.F. Frosch et al., Phys. Rev. {\bf 160} (1968) 874

\bibitem {Ed} E.L. Tomusiak, W. Leidemann and H.M. Hofmann, Phys.
 Rev. {\bf C52} (1995) 1963
\bibitem {Deto} R. Detomo et al., unpublished data from Ohio State
University; private communication from T. Donoghue (1978)
\bibitem {Erlangen} R. Kankowsky, J.C. Fritz, K. Kilian, A. Neufert
 and D. Fick, Nucl. Phys. {\bf A263} (1976) 29
\bibitem {LA} R.F. Haglund, Jr., R.E. Brown, N. Jarmie, G.G. Ohlsen,
 P.A. Schmelzbach and D. Fick, Phys. Lett. {\bf 79B} (1978) 35
\bibitem {Perry} J.E. Perry, Jr. et al. (unpublished), reported by
J.D. Seagrave in Proceedings of the Conference on Nuclear Forces
 and the Few Nucleon Problem, London (1959)  
\bibitem {Klages} B. Haesner, Dissertation, Kernforschungszentrum
 Karlsruhe (August 1982); private communication from H. Klages (1995)
\bibitem {Drosg} M. Drosg, D.K. McDaniels, J.C. Hopkins, J.D. Seagrave,
 R.H.  Sherman and E.C. Kerr, Phys. Rev. {\bf C9} (1974) 179
\bibitem {Drigo} L. Drigo, G. Tornielli and G. Zannoni, Ann. Phys.  {\bf 7}
(1982) 408
\bibitem {Lisow} P.W. Lisowski, C.T. Rhea and R.L. Walter, Nucl.
 Phys. {\bf A259} (1976) 61
\bibitem {grueb} W. Gr\"uebler, V. K\"onig, P.A. Schmelzbach, R.  Risler,
R.E. White and P. Marmier, Nucl. Phys. {\bf A193} (1972) 129
\bibitem {schulte} R.L. Schulte, M. Cosack, A.W. Obst and J.L Weil, Nucl.
Phys. {\bf A192} (1972) 609
\bibitem {dries} L.J. Dries, H.W. Clark, R. Detomo, Jr. and T.R.
Donoghue, Phys. Lett. {\bf 80B} (1979) 176
\bibitem {ddc} J.E. Brolley, Jr., T.M. Putnam, L. Rosen and L. Stewart,
Phys. Rev. {\bf 117} (1960) 1307
\bibitem {grueb73} W. Gr\"uebler, V. K\"onig, R. Risler, P.A. Schmelzbach,
R.E. White and P. Marmier, Nucl. Phys. {\bf A193} (1972) 149
\bibitem {bonntensor} R. Machleidt, Adv. Nucl. Phys. {\bf 19} (1991) 230
\bibitem {av14} R.B. Wiringa, R.A. Smith and T.L. Ainsworth, Phys. Rev. C
{\bf 29} (1984) 1207
\end{thebibliography}
\end{document}